\pdfoutput=1
\documentclass[12pt,letterpaper,floatfix]{appolb}
\usepackage{amsmath}
\usepackage{graphicx}
\usepackage{amssymb}
\usepackage{pslatex}
\usepackage{epsfig}
\usepackage{euscript}
\usepackage{fancybox}
\usepackage{color}

\def\mathswitchr#1{\relax\ifmmode{\mathrm{#1}}\else$\mathrm{#1}$\fi}

\newcommand {\KKMC}{\hbox{${\cal KK}$}\ MC}

\newcommand {\pslash}{\hbox{$\not\hbox{\kern-2.3pt $p$}$}}

\usepackage{cite}

\usepackage{epic}

\def\alf1{ {\alpha\over\pi} }
\def\rQCED{{\rm QCED}}

\begin{document}
\begin{titlepage}
\begin{flushright}
{\bf BU-HEPP-14-01}\\
{\bf Apr., 2014}\\
\end{flushright}
\vspace{0.05cm}
 
\begin{center}
{\Large Current Status of LHC Physics: Precision Theory View}
\end{center}

\vspace{2mm}
\begin{center}
{\bf   B.F.L. Ward}\\
\vspace{2mm}
{\em Department of Physics,}\\
{\em Baylor University, Waco, Texas, USA}\\
\end{center}

\vspace{5mm}
\begin{center}
{\bf   Abstract}
\end{center}
We discuss the current status of LHC physics from the perspective of precision theory predictions for the attendant QCD and EW higher order corrections.
We focus on the interplay between the available data and the predictions
for these data viewed in the context of the establishment of baselines
for what is needed to exploit fully the discovery potential of the existing LHC data and the data expected by the end of the second LHC run(i.e., 300fb$^{-1}$). We conclude that significant improvement in the currently used theoretical predictions will be mandatory. Possible strategies to achieve such improvement are indicated.
\\
\vskip 3mm
\centerline{Invited talk presented at the 2014 Epiphany Conference, Krakow, Poland}
\vskip 16mm
\vspace{10mm}
\renewcommand{\baselinestretch}{0.1}
\footnoterule
\noindent
{\footnotesize
}

\end{titlepage}
 
\baselineskip=11pt 
\def\Kmax{K_{\rm max}}\def\ieps{{i\epsilon}}\def\rQCD{{\rm QCD}}
\renewcommand{\theequation}{\arabic{equation}}
\font\fortssbx=cmssbx10 scaled \magstep2
\renewcommand\thepage{}
\parskip.1truein\parindent=20pt\pagenumbering{arabic}\par

\section{\bf Introduction}\label{intro}\par
As the LHC has operated successfully for a considerable period and has discovered~\cite{atlas-cms-2012} the long sought BEH~\cite{BEH}, paving the way for the 2013 Nobel Prize in Physics for Profs. F. Englert and P. Higgs, and 
as the LHC is now in its Long Shutdown \#1 as it prepares to move to higher energies, it is appropriate to asses the status of the physics purview of the LHC
as it currently stands. This is our objective in what follows here.\par
With such an objective, it is appropriate as well to assume a particular vantage point, as one can well imagine a number of such. We will take the perspective
of the theory of precision LHC physics, by which we mean
predictions for LHC processes at the total precision tag of $1\%$ or better.
It is appropriate for any discussion from the perspective of precision theory
for the status of physics purview of the LHC  
to set the attendant framework by recalling, at least in generic terms, 
why we {\em still} need the LHC in the first place. In the following discussion we shall begin with such recapitulation. In this way, the entirety 
of the effort required to realize and to extend the current purview
of the physics for the LHC in a practical 
way can be more properly assessed.\par
Thus, in view of the discovery of the BEH boson we ask, ``Why do we {\em still} need the LHC?'' 
Many answers can be found in the 
original justifications for the colliding beam device and its detectors in Refs.~\cite{lhc,atlas,cms,lhcb,alice}. We will call attention to a particular
snap shot
of the latter discussions with some eye toward the requirements of precision 
theory for LHC physics in view of the discovery of the BEH boson. More precisely, the LHC physics program still remains as a crucial step toward resolving fundamental outstanding issues in elementary particle physics:
the big and little hierarchy problems,
the number of families,
the origin of Lagrangian fermion (and gauge boson) masses,
baryon stability,
the union of quantum mechanics and general theory of relativity,
the origin of CP violation,
the origin of the cosmological constant $\Lambda$,
dark matter,~
$\cdots$ .
The discovery of the BEH boson serves to help us refocus
much of the considerable theory effort 
that has been invested in the ``New Physics'' (NP) 
that would still seem to be needed to solve all of these outstanding issues, 
that is to say, in the physics beyond the Standard Model 't Hooft-Veltman 
renormalized
Glashow-Salam-Weinberg EW $\otimes$ Gross-Wilczek-Politzer QCD theory that
seems to describe the quantum loop corrections in the measurements of 
electroweak and strong interactions at the shortest distances 
so far achieved in laboratory-based experiments.\par
We may still mention that superstring theory~\cite{gsw,pol} solves everything 
in principle but has trouble in practice: for example it has more 
than $10^{500}$
candidate solutions for the vacuum state~\cite{stringvac}. The ideas in 
superstring theory have helped to motivate many so-called string inspired 
models of NP such as~\cite{bsm} string-inspired GUTs, large extra dimensions, Kaluza-Klein excitations, ... . 
We would list supersymmetric extensions of the SM, such as the 
MSSM, the CMSSM and the more recent pMSSM~\cite{bsm}, 
as separate proposals from 
superstring motivated ideas, as historically this was the case. Modern 
approaches to the dynamical EW symmetry  breaking 
(technicolor) such as little Brout-Englert-Higgs models~\cite{bsm} 
obtain as well. The list is quite long and LHC, especially now that
it has found the BEH boson, will clearly help us shorten it.\par
One of the most provocative ideas continues to be 
the one which some superstring 
theorists~\cite{stringvac} invoke to solve the problem of the large number of 
candidate superstring vacua: the anthropic principle, by which the solution 
is the one that allows us to be in the state in which we find ourselves. 
In the view of some~\cite{susskd}, this would be the end 
of reductionist physics as we 
now know it. With the discovery of the BEH boson, has the chance that the LHC 
can even settle this discussion decreased? Perhaps not.\par
Recently, even newer paradigms are emerging which would foretell the need for new accelerated beam devices. In Ref.~\cite{leguts}, it is shown that
GUT's exist in which there are three or more heavy families such that
SM quarks are paired with new heavy leptons and SM leptons are paired with new
heavy quarks with a GUT scale $M_{GUT}\sim 100$TeV, so that it would be accessible to the VLHC device as discussed in Ref.~\cite{vlhc} as well as at the
FCC-ee device under discussion now at CERN~\cite{fcc}. The proton is stable because all the leptons to which it could decay are too heavy for the decay to occur.
Indeed, many of the new states in such scenarios may already be accessible at the LHC. New ideas are also needed as a remedy for the superficially bad UV
behavior of quantum gravity if one does not use  string theory for the unification of the EW and QCD theories with quantum gravity and these new ideas
also foretell the need for new accelerated beam devices as well as the possible existence of new signatures at the LHC. More specifically, more progress has been made recently on solving the problem of the UV behavior of quantum gravity
in the context of local Lagrangian field theory methods~\cite{reuter-picac-litm,rqg,kreimer}. Indeed, following the suggestion by Weinberg~\cite{wein1} that quantum gravity might have a non-trivial UV fixed point, with a finite dimensional critical surface
in the UV limit, so that it would be asymptotically safe with an S-matrix
that depends on only a finite number of observable parameters, 
in 
Refs.~\cite{reuter-picac-litm} 
strong evidence has been calculated
using Wilsonian~\cite{kgw} field-space exact renormalization group 
methods to support
Weinberg's asymptotic safety hypothesis for the Einstein-Hilbert theory
\footnote{In addition, more evidence for Weinberg's asymptotic safety behavior has been calculated using causal dynamical triangulated lattice methods in Ref.~\cite{ambj}.}
\footnote{We also note that the model in Ref.~\cite{horva} realizes many aspects
of the effective field theory implied by the anomalous dimension of 2 at the
UV-fixed point but it does so at the expense of violating Lorentz invariance.}.
In a parallel but independent development~\cite{rqg}, we have shown~\cite{bw2i} that the extension of the amplitude-based, exact resummation theory of Ref.~\cite{yfs} to the Einstein-Hilbert theory leads to analogous UV-fixed-point behavior for the dimensionless gravitational and cosmological constants with the bonus that the resummed theory is actually UV finite when expanded in the resummed propagators and vertices to any finite order in the respective improved loop expansion.
We designate the resummed theory as resummed quantum gravity. 
We emphasize that there is no known inconsistency between our analysis
and those of the Refs.~\cite{reuter-picac-litm,ambj} or
the leg renormalizability arguments in Ref.~\cite{kreimer}.
In Ref.~\cite{bw-lambda} we use resummed quantum gravity in the context
of the Planck scale cosmology of Refs.~\cite{reuter1,reuter2}\footnote{The attendant choice of the scale $k\sim 1/t$ used in Refs.~\cite{reuter1,reuter2} was also proposed in Ref.~\cite{sola1}.}, which is based on the
approach of Refs.~\cite{reuter-picac-litm}, to predict the value of the cosmological constant such that $\rho_\Lambda\cong(2.40\times 10^{-3}\text{\rm eV})^4$. This latter result is close enough
to the observed value~\cite{cosm1,pdg2008}, $((2.37\pm 0.05)\times 10^{-3}\text{rm eV})^4$, that it would require some~\cite{ravi} susy GUT scenarios to have
flipped copies of their EW scale multiplets such that these copies would also be
accessible at VLHC and the newly discussed FCC devices.
We may also note that new approaches to dark matter 
such as Higgsogenesis~\cite{servant}
suggest new particles in the 100 TeV regime could obtain, again
in reach of the VLHC and FCC devices. Finally, to end our illustration
of the many new paradigms in the literature\footnote{We apologize right now for all the new ideas we have not mentioned for reasons of space.}, 
we note that in Ref.~\cite{giud-dvli}
the UV limit of theories such as quantum gravity is solved by the dynamical generation of non-perturbative large distance excitations called classicalons, which provide the necessary damping of the naively divergent UV behavior. When discussed in general terms, possible new signatures for the LHC, VLHC and FCC obtain~\cite{giud-dvli}.\par
Evidently, the new paradigms,  
admittedly only illustrated here in a limited way to set the stage of our 
discussion, must be taken seriously in analyzing the new LHC data. 
We continue to stress that 
we must be able to distinguish higher order SM processes 
from New Physics and that we must be able to probe New Physics precisely to 
distinguish among different New Physics scenarios. These requirements 
necessitate the 
era of precision theory for the LHC and justify it as a valid perspective from which to view the current status of LHC physics.\par
\begin{figure}[ht]
\begin{center}
\includegraphics[width=80mm]{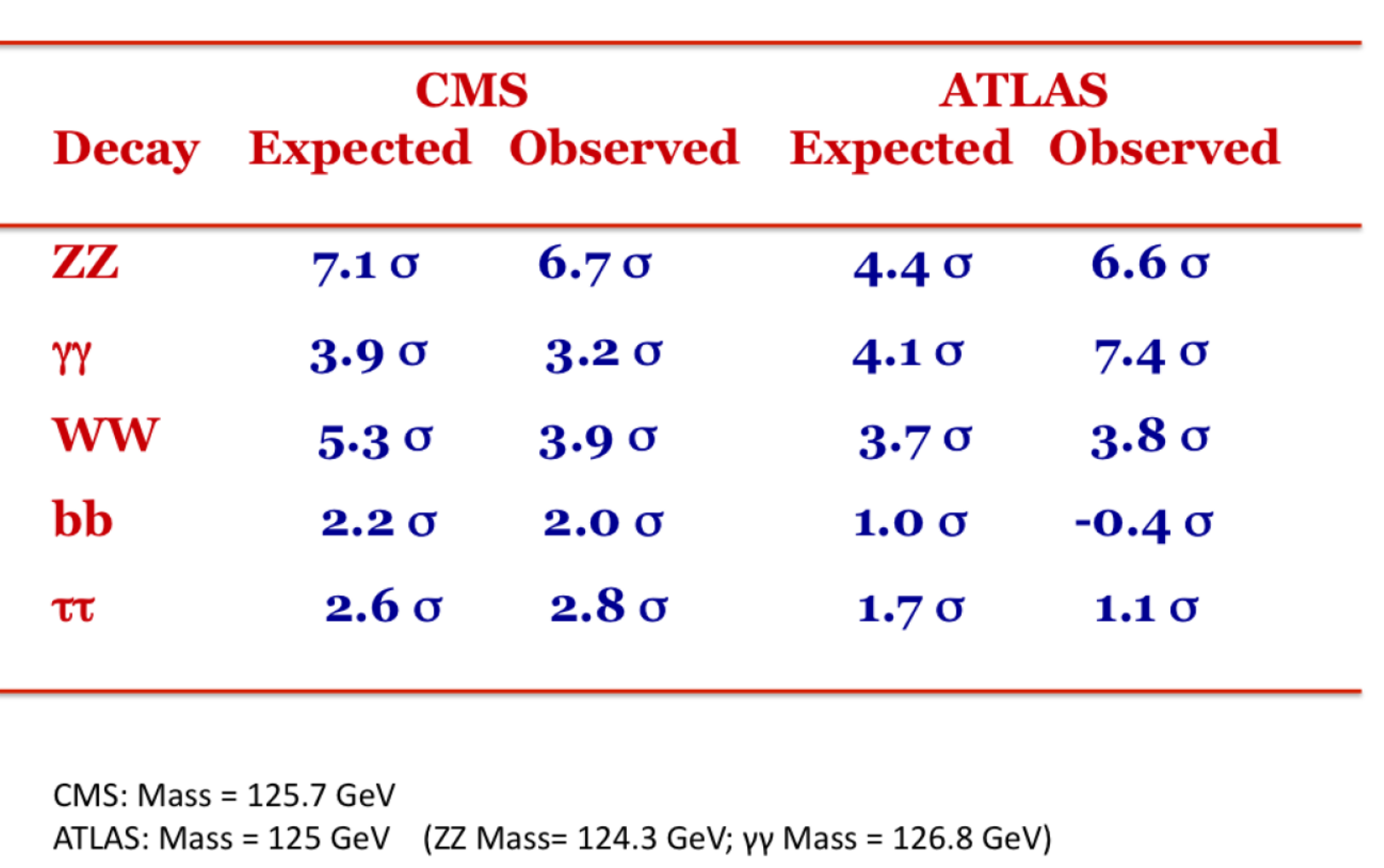}
\end{center}
\caption{\baselineskip=11pt  Results on BEH boson observation at the 
LHC as reviewed in Ref.~\cite{deRk-lp13}.}
\label{fig1-deRk}
\end{figure}

To give an overview of our discussion, we call attention in Fig.~\ref{fig1-deRk}
to the summary of the observations of the BEH boson by ATLAS and CMS as reported
in Ref.~\cite{deRk-lp13}. 
Given the greater than $9\sigma$ significance level in each
collaboration's observation, we proceed in what follows by evaluating what
will be needed from the standpoint of precision theory to make the corresponding improvement in our knowledge of the BEH boson and its properties and of the
limits on its possible extensions/generalizations(NP physics) given the expected improvements in the integrated luminosity and energies during the future planned running of the LHC. We start in the next Section with the BSM Higgs scenarios but we discuss as well the precision BEH boson studies from this latter perspective. We then address in Section 3 what issues obtain in developing the theory improvements so implied.
We sum up in our Conclusions.\par
\section{BSM and Precision BEH Scenarios}
For a more detailed discussion of BSM scenarios relative to the current status of LHC physics, see Ref.~\cite{pokorski-epi14}. Here, we use a few representative examples to illustrate the attendant view from the perspective of precision theory.\par
\begin{figure}[h]
\begin{center}
\includegraphics[width=80mm]{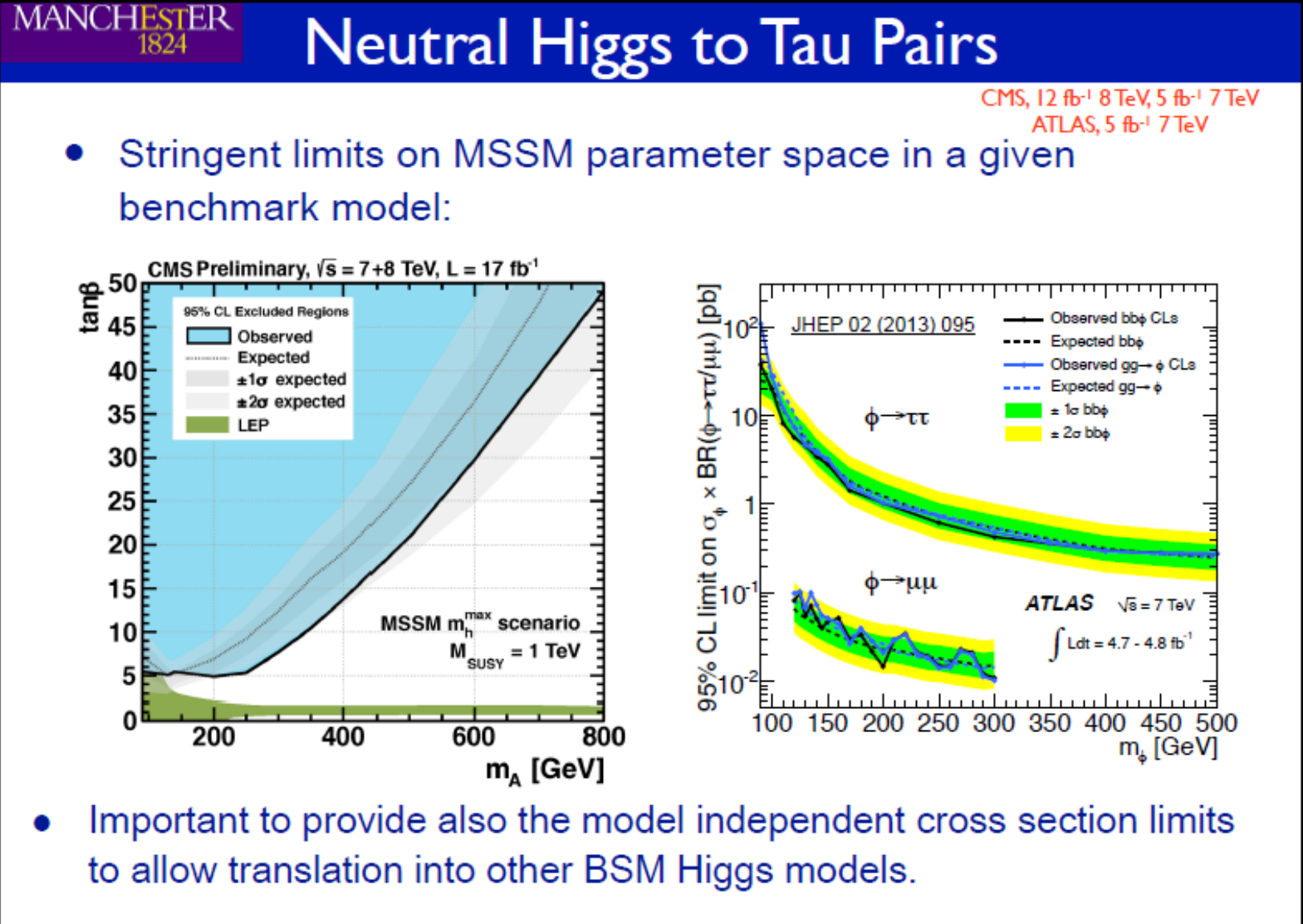}
\end{center}
\caption{\baselineskip=11pt  Stringent limits are put on a benchmark (MSSM) susy model's parameter space from neutral Higgs 
decays to tau pairs as constrained by data at the 
LHC as reviewed in Ref.~\cite{own-lp13}.}
\label{fig2-own}
\end{figure}
\begin{figure}[h]
\begin{center}
\includegraphics[width=80mm]{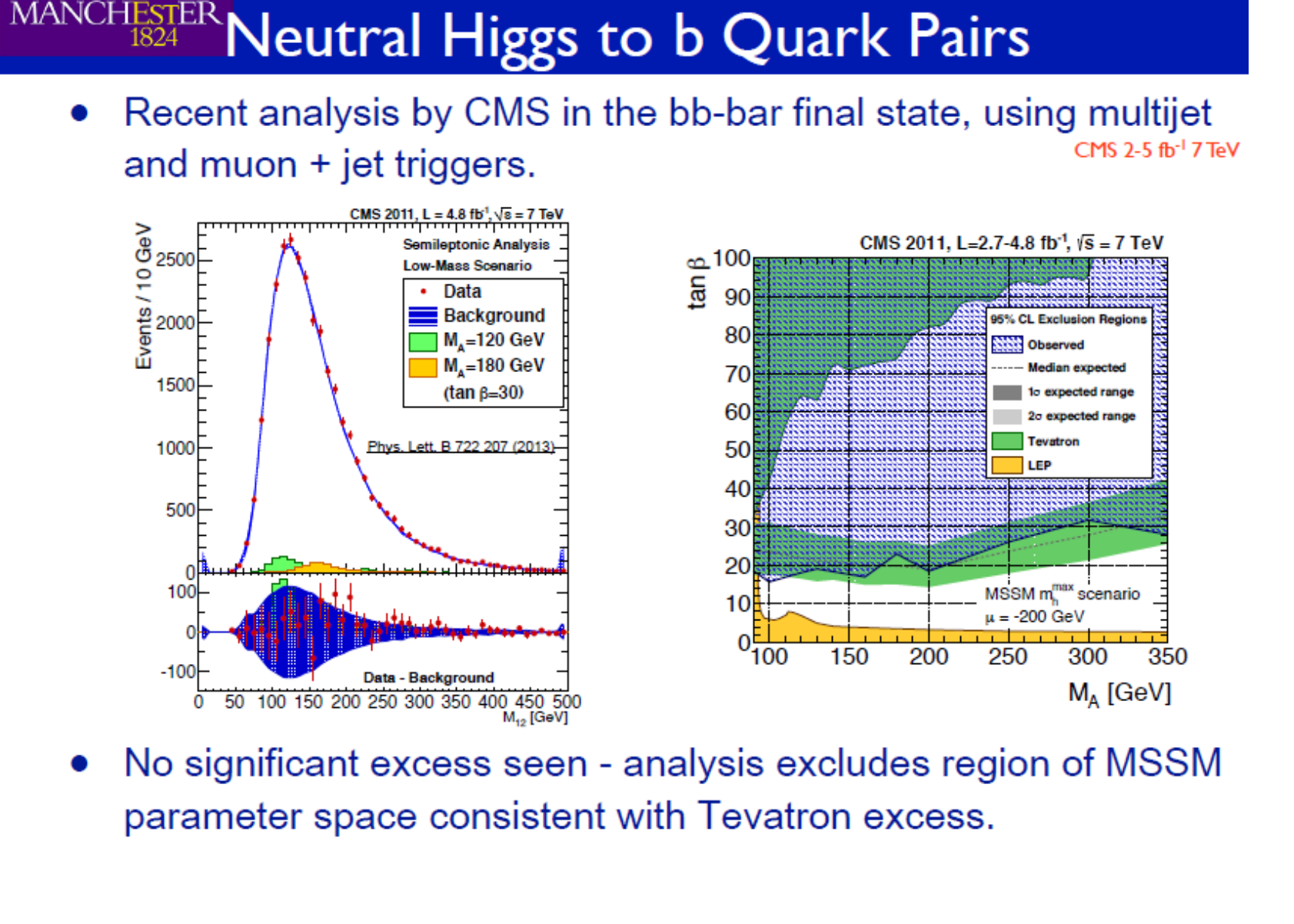}
\end{center}
\caption{\baselineskip=11pt  Limits on MSSM parameter space from neutral Higgs 
decays to b quark pairs as constrained by data at the 
LHC as reviewed in Ref.~\cite{own-lp13}.}
\label{fig3-own}
\end{figure}
\begin{figure}[h]
\begin{center}
\includegraphics[width=170mm]{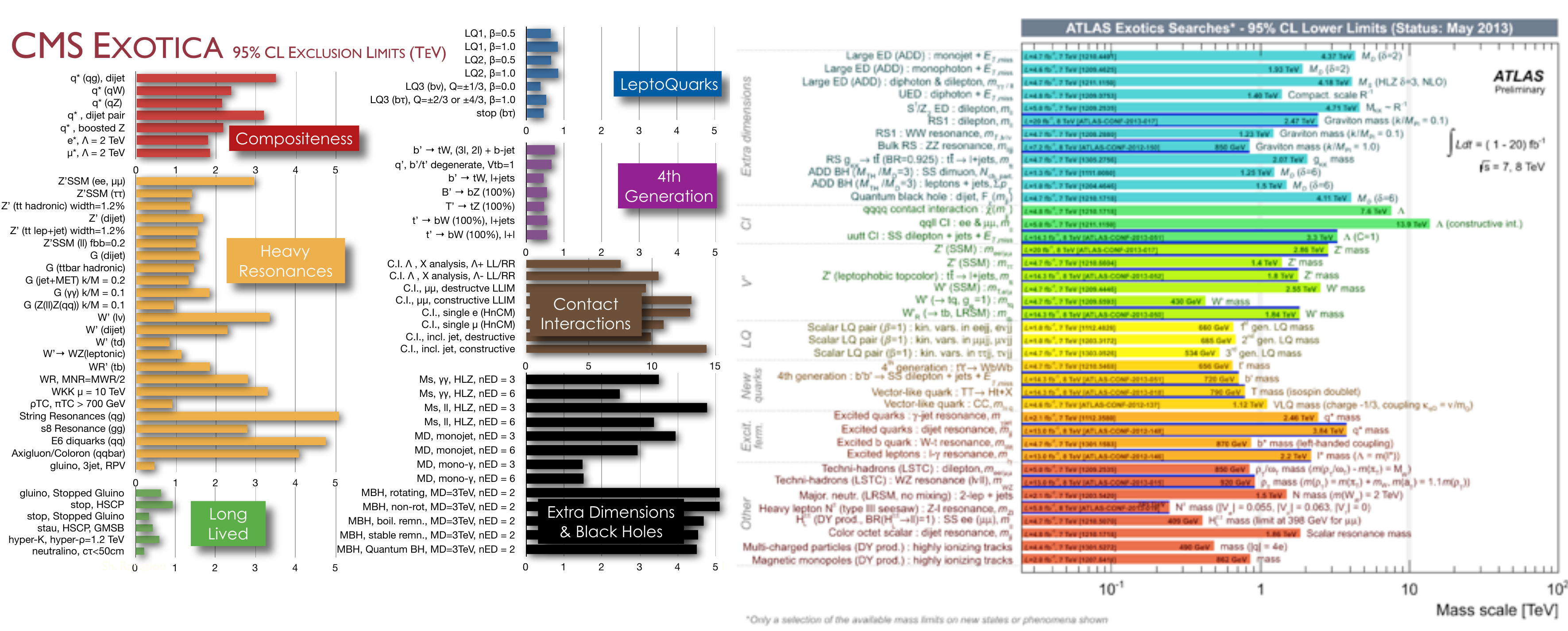}
\end{center}
\caption{\baselineskip=11pt  Limits on general BSM models from LHC data
as reviewed in Ref.~\cite{rapp-lp13}.}
\label{fig4-rappoccio}
\end{figure}

In Ref.~\cite{own-lp13}, the constraints from LHC data on new neutral Higgs particles in the MSSM in their decays 
are used to set the limits on the
respective MSSM parameter space as illustrated 
in Fig.~\ref{fig2-own}(tau pairs)
and Fig.~\ref{fig3-own}(b quark pairs).
In Ref.~\cite{rapp-lp13} the general situation on the constraints of LHC data on BSM scenarios is reviewed and we show in Fig.~\ref{fig4-rappoccio} a summary
of the respective results as presented therein.
When we look at the results just discussed, we ask, ``How are these results obtained?'' The important answer is that, in all cases, they obtain from
comparison of theoretical predictions with experimental data. This allows 
a future for LHC physics that entails precision studies of the properties of the BEH boson and precision studies of the possibilities for BSM physics, both of which open possibilities for fundamental discoveries at the LHC. We now develop
this view of the future of LHC physics based on what has been achieved at this writing.\par
\begin{figure}[h]
\begin{center}
\includegraphics[width=80mm]{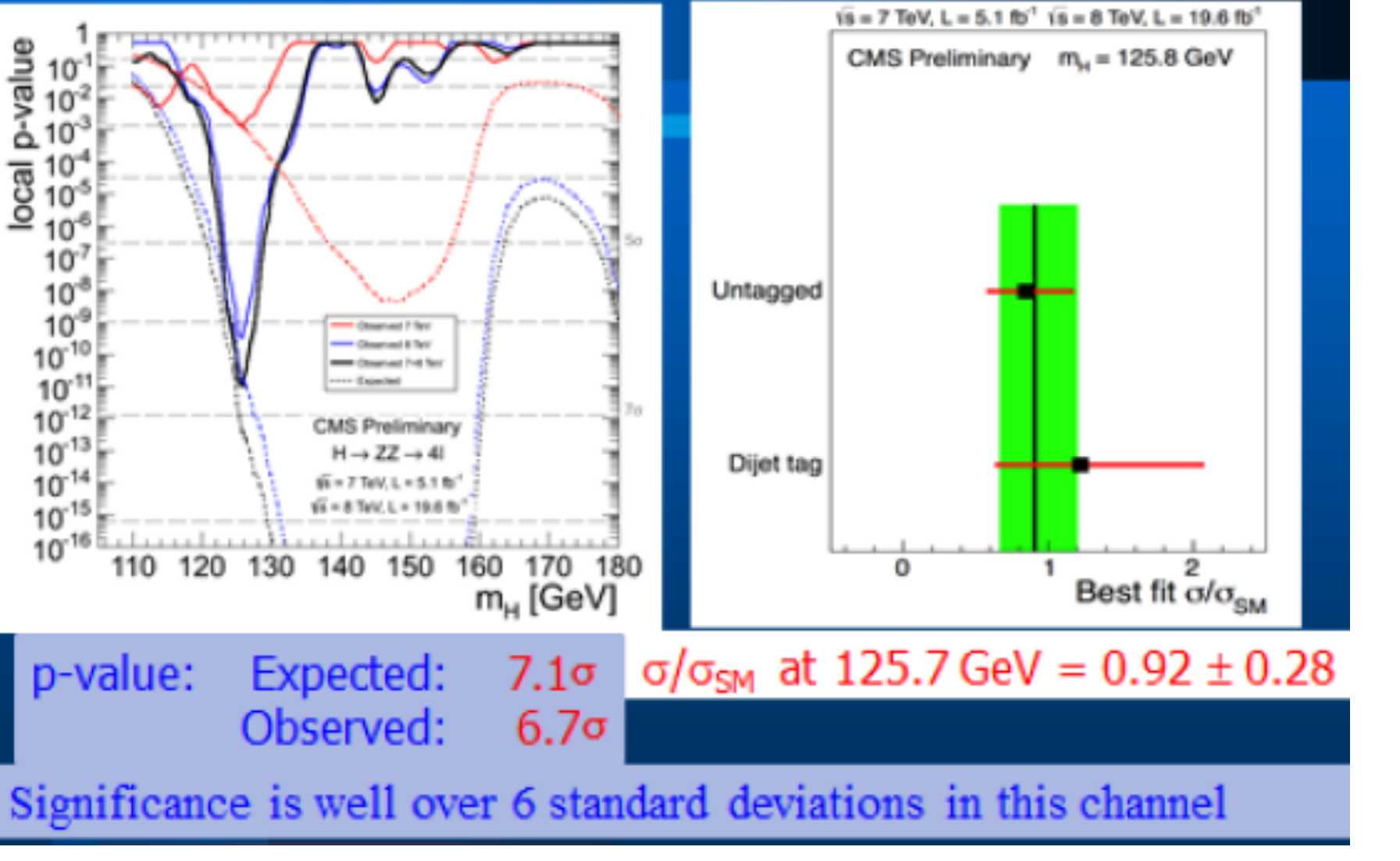}
\end{center}
\caption{\baselineskip=11pt Measurements of $\mu=\sigma/\sigma_{\text{SM}}$ 
for the channel $H\rightarrow ZZ\rightarrow 4\ell$ from CMS as 
reviewed in Ref.~\cite{deRk-lp13}.}
\label{fig5-deRk}
\end{figure}
\begin{figure}[h]
\begin{center}
\includegraphics[width=80mm]{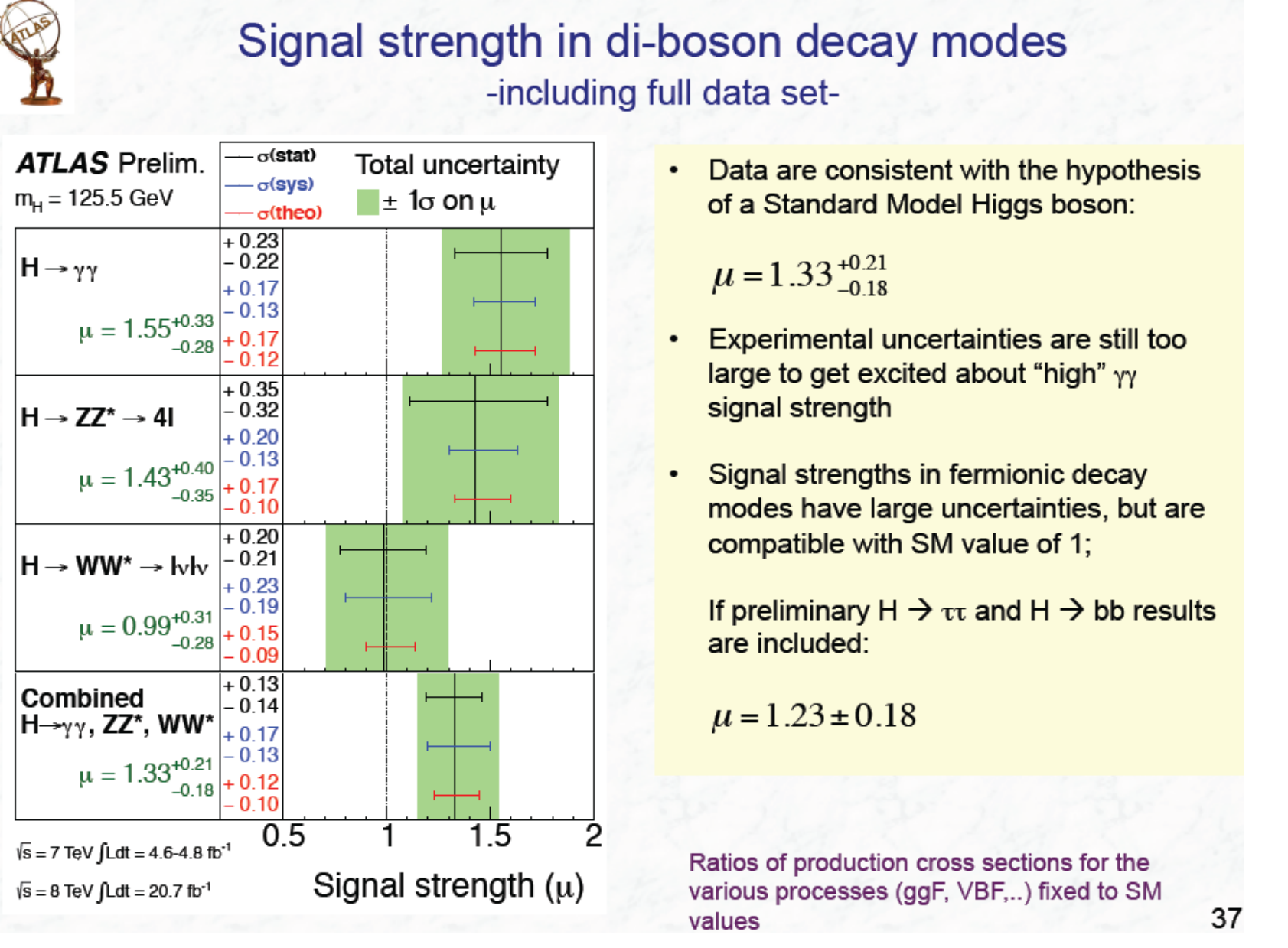}
\end{center}
\caption{\baselineskip=11pt Measurements of signal strengths in diboson decay modes for the 
Higgs from ATLAS as reviewed in Ref.~\cite{jcbs-lp13}.}
\label{fig6-jcbs}
\end{figure}

Starting first with the implications for precision BEH studies, we show respectively in Fig.~\ref{fig5-deRk}~\cite{deRk-lp13} and in 
Fig.~\ref{fig6-jcbs}~\cite{jcbs-lp13} the current constraints
from LHC data on $\sigma/\sigma_{\text{SM}}\equiv\mu$ for the $H\rightarrow ZZ \rightarrow 4\ell$ H production as measured by CMS and a similar set of constraints from ATLAS which entails the same channel as well as two others, the $H\rightarrow \gamma\gamma$ and $H\rightarrow WW \rightarrow \ell\nu\ell\nu$ channels.
Here, $\sigma(\sigma_{\text{SM}})$ is the cross section for the respective channels as observed(as predicted in the SM) respectively. 
What we see in these results, which are illustrative of the generic situation in many current BEH studies, is that we have a $10-12\%$ value for the theoretical uncertainty $\Delta\mu_{\text{th}}$ on $\mu$ versus the respective experimental uncertainty $\Delta\mu_{\text{expt}}\sim 20\%$. This is sufficient currently, but, as we see from Ref.~\cite{wells-lp13}, the expectation is that LHC will have delivered $300fb^{-1}$ by 2021 and to achieve the analogous results we would need by that time  
$\Delta\mu_{\text{th}}\lesssim 3.2\%$. This requires precision SM theory with a provable precision tag, in analogy with what was done for LEP -- see Refs.~\cite{lep-prec}.\par
\begin{figure}[h]
\begin{center}
\includegraphics[width=80mm]{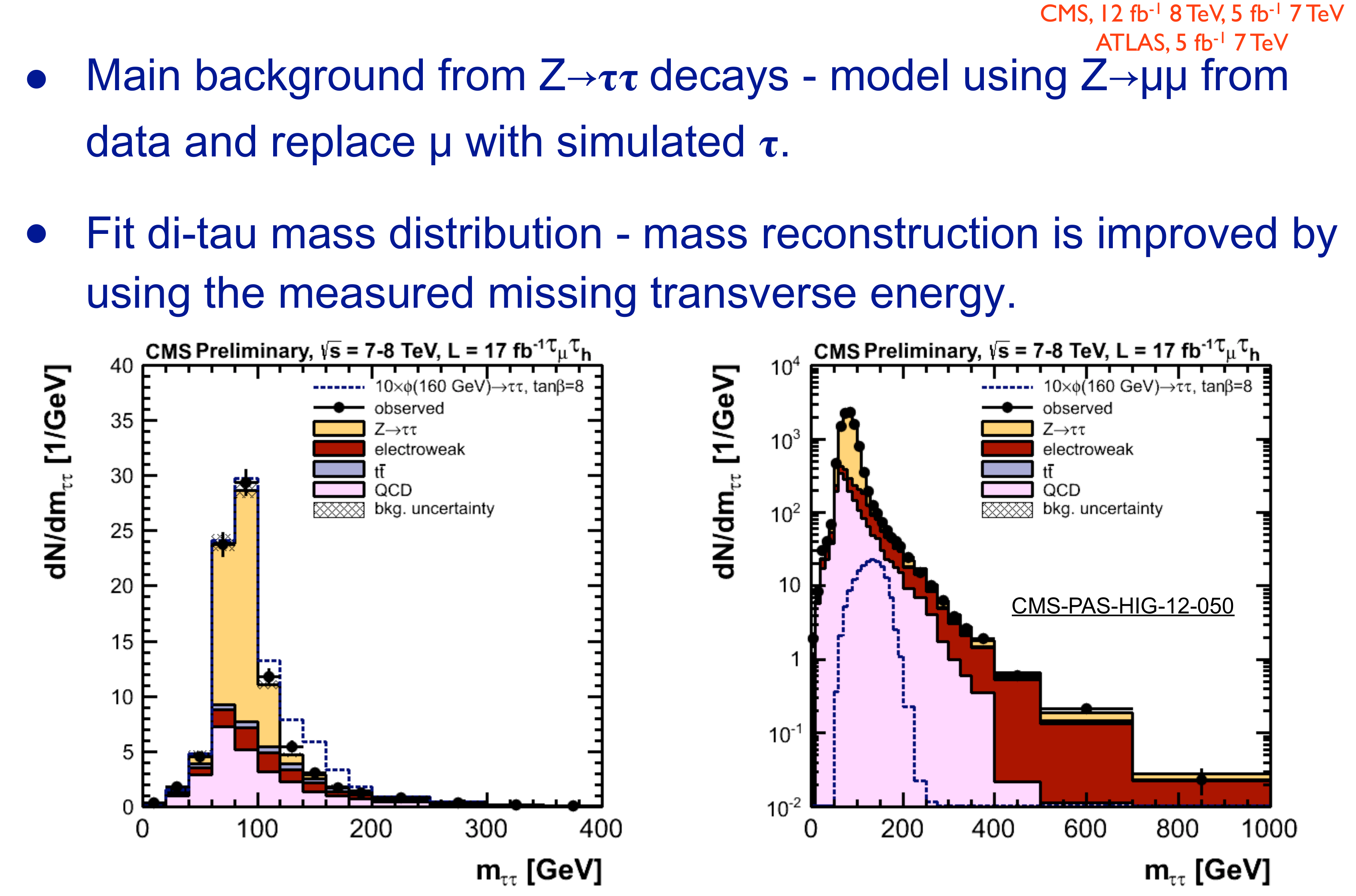}
\end{center}
\caption{\baselineskip=11pt Measurements of di-tau mass spectrum in search for
neutral Higgs decaying to $\tau\bar{\tau}$ at LHC as 
reviewed in Ref.~\cite{own-lp13}. The signal of a 160 GeV MSSM possible realization is shown for illustration for $\tan\beta=8$.}
\label{fig7-own}
\end{figure}

  
A similar situation obtains for BSM studies. To illustrate this point
we show in 
Fig.~\ref{fig7-own} together with what we showed in Fig.~\ref{fig2-own}
the studies of a neutral Higgs decaying to $\tau\bar{\tau}$ from CMS and ATLAS as reviewed in Ref.~\cite{own-lp13}. 
These results feature the experimental precision of $\Delta\sigma_{\text{expt}}\cong 3.4\%$ to be compared with the theoretical precision $\Delta\sigma_{\text{th}}\cong 2.8\%$. With 300$fb^{-1}$ of data, to get the analogous return on the data analysis we need $\Delta\sigma_{\text{th}}\lesssim 1\%$. This requires precision theory for both QCD and EW corrections for LHC physics.\par
\begin{figure}[h]
\begin{center}
\includegraphics[width=80mm]{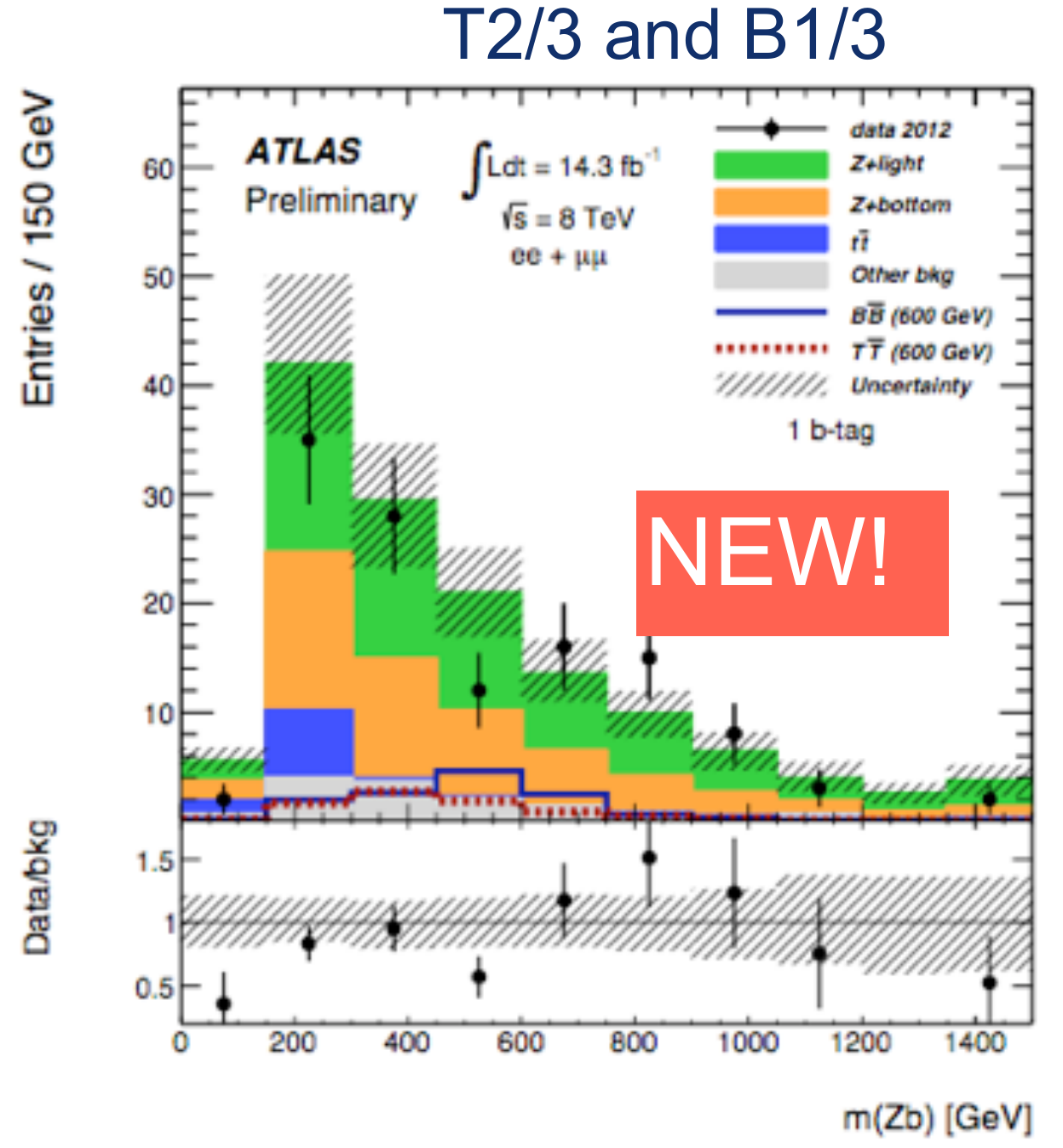}
\end{center}
\caption{\baselineskip=11pt Data from the search for $T^{2/3},\; B^{1/3}$ in ATLAS as reviewed in Ref.~\cite{rapp-lp13}.}
\label{fig9-rapp}
\end{figure}

Continuing in this way, we show in Fig.~\ref{fig9-rapp} the data from the ATLAS search for new quark partners $T^{2/3},\; B^{1/3}$ as reviewed in Ref.~\cite{rapp-lp13}. The current precisions, which yield the mass limits for 0.7TeV for generic branching ratios~\cite{rapp-lp13}, are $\Delta\sigma_{\text{expt}}\cong 10\%$ and $\Delta\sigma_{\text{th}}\cong 10\%$ and, for 300$fb^{-1}$ one would need $\Delta\sigma_{\text{th}}\lesssim 2\%$. Again, this requires precision theory for the QCD and EW corrections to LHC physics.
\par
\begin{figure}[h]
\begin{center}
\includegraphics[width=80mm]{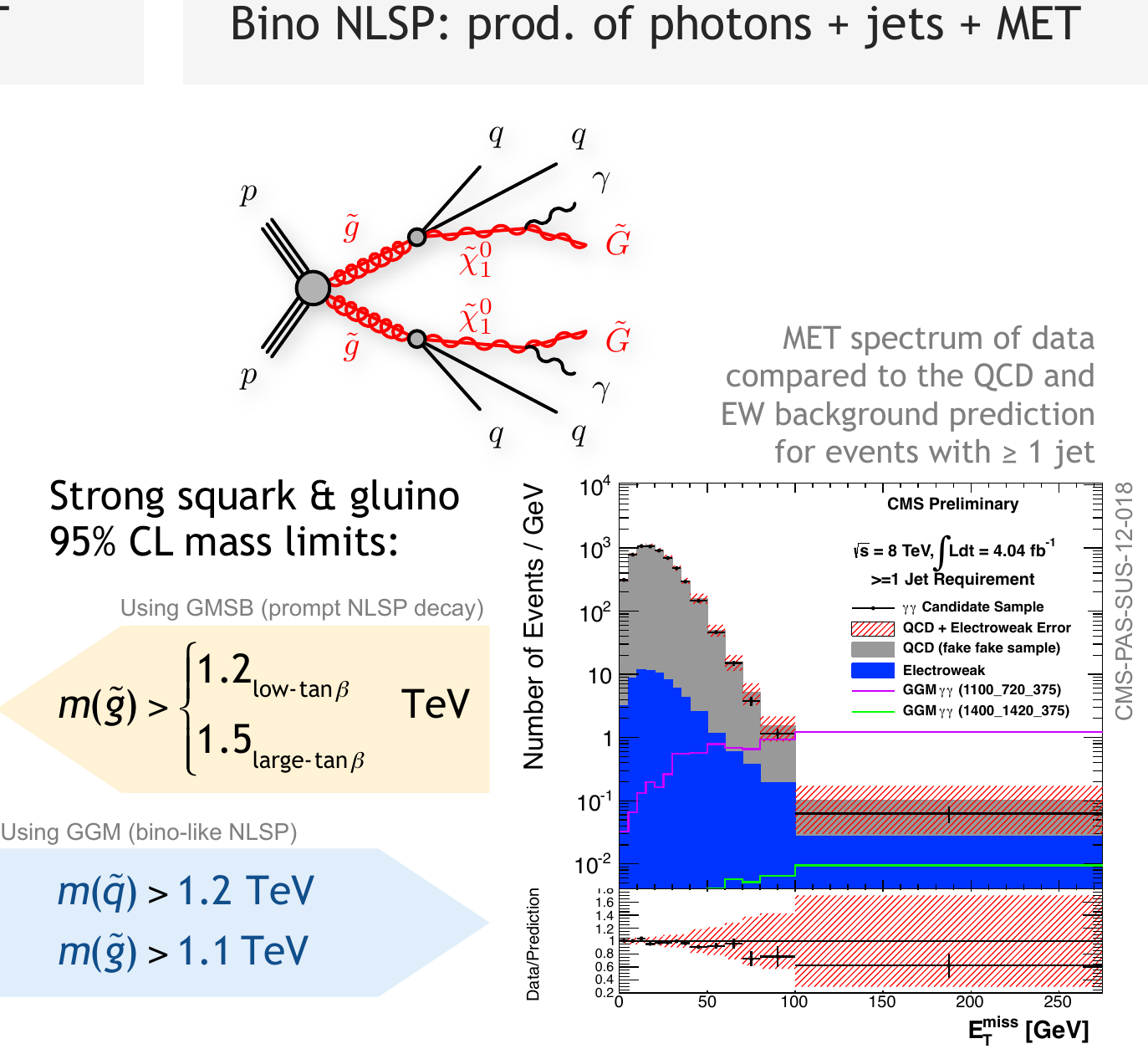}
\end{center}
\caption{\baselineskip=11pt Data from the search for the bino NLSP with
production of $\gamma$'s + jets + MET from CMS as reviewed in Ref.~\cite{hoekr-lp13}.}
\label{fig10-hoekr}
\end{figure}

The generic searches for susy have a similar message, as we see in Fig.~\ref{fig10-hoekr} where we feature results reviewed in Ref.~\cite{hoekr-lp13} for LHC susy searches using the MET spectrum for events with $\ge 1$ jet.
Currently, for $E^{MISS}_T\gtrsim 75$GeV we have the experimental error $\Delta\sigma_{\text{expt}}\sim 25\%$ and the theoretical error $\Delta\sigma_{\text{th}}\sim 40\%$, which imply the limits, using GGM (bino-like NLSP), $m(\tilde{q}) > 1.2$ TeV,\; $m(\tilde{g})> 1.1$ TeV for the respective squark and gluino mass limits, for example. With 300$fb^{-1}$ of data, to get a similar return on the data, we need $\Delta\sigma_{\text{th}}\lesssim 10\%$, which again motivates precision theory for the QCD and EW corrections.\par
\begin{figure}[h]
\begin{center}
\includegraphics[width=80mm]{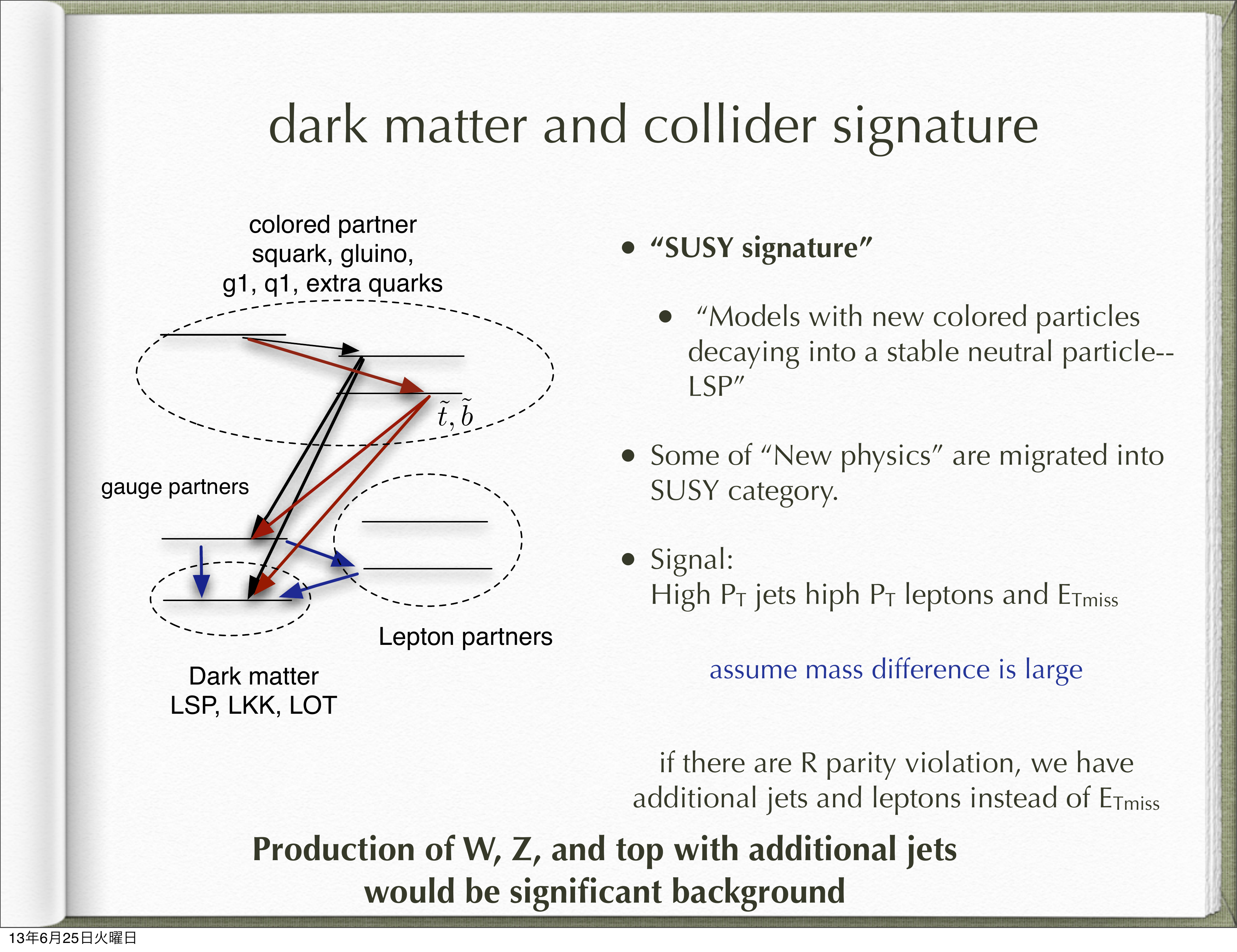}
\end{center}
\caption{\baselineskip=11pt Generic New Physics scenarios as reviewed in Ref.~\cite{njri-lp13} .}
\label{fig11-njri}
\end{figure}
\begin{figure}[h]
\begin{center}
\includegraphics[width=80mm]{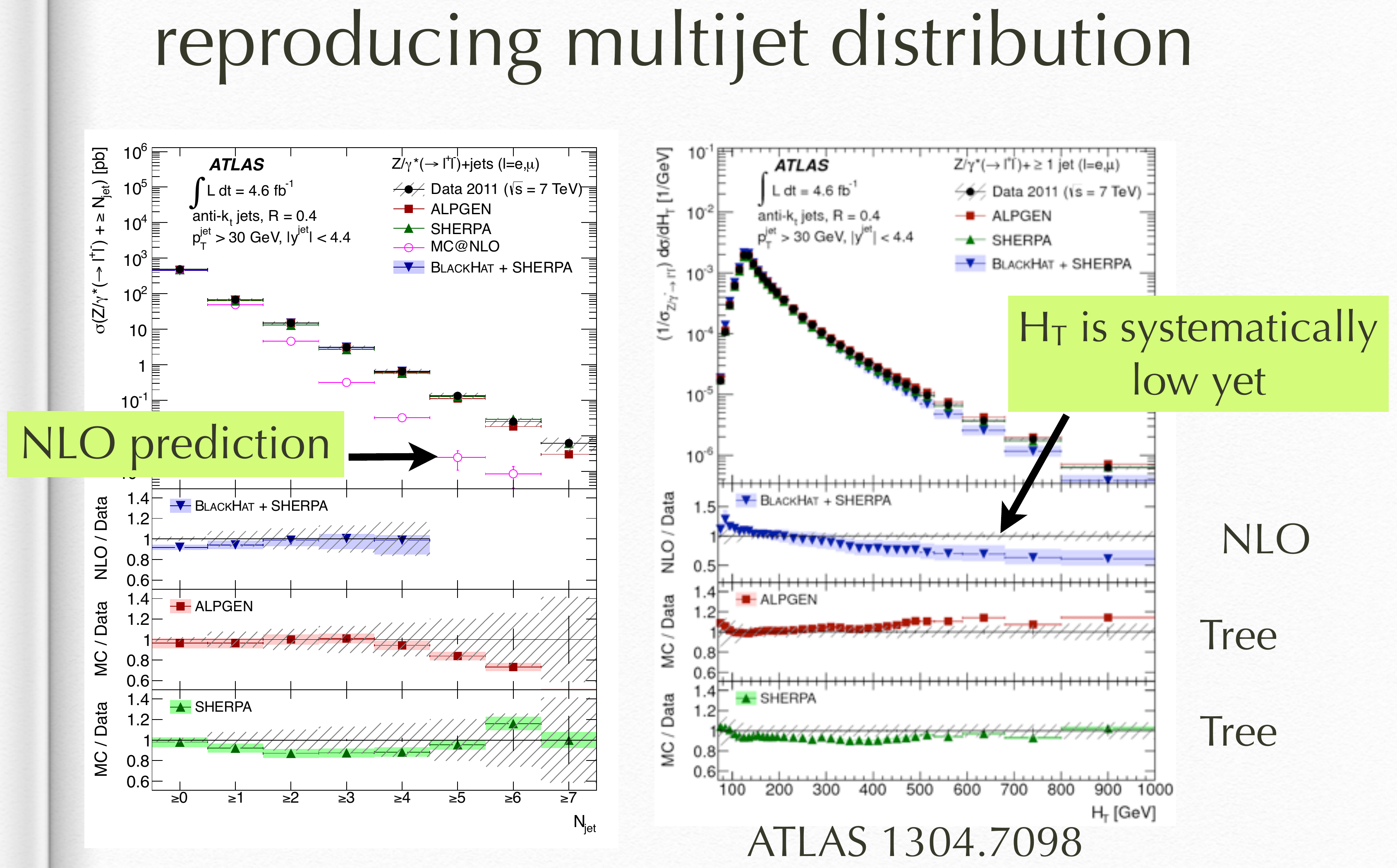}
\end{center}
\caption{\baselineskip=11pt Various available background estimates as reviewed in Ref.~\cite{njri-lp13}.}
\label{fig12-njri}
\end{figure}
\begin{figure}[h]
\begin{center}
\includegraphics[width=100mm]{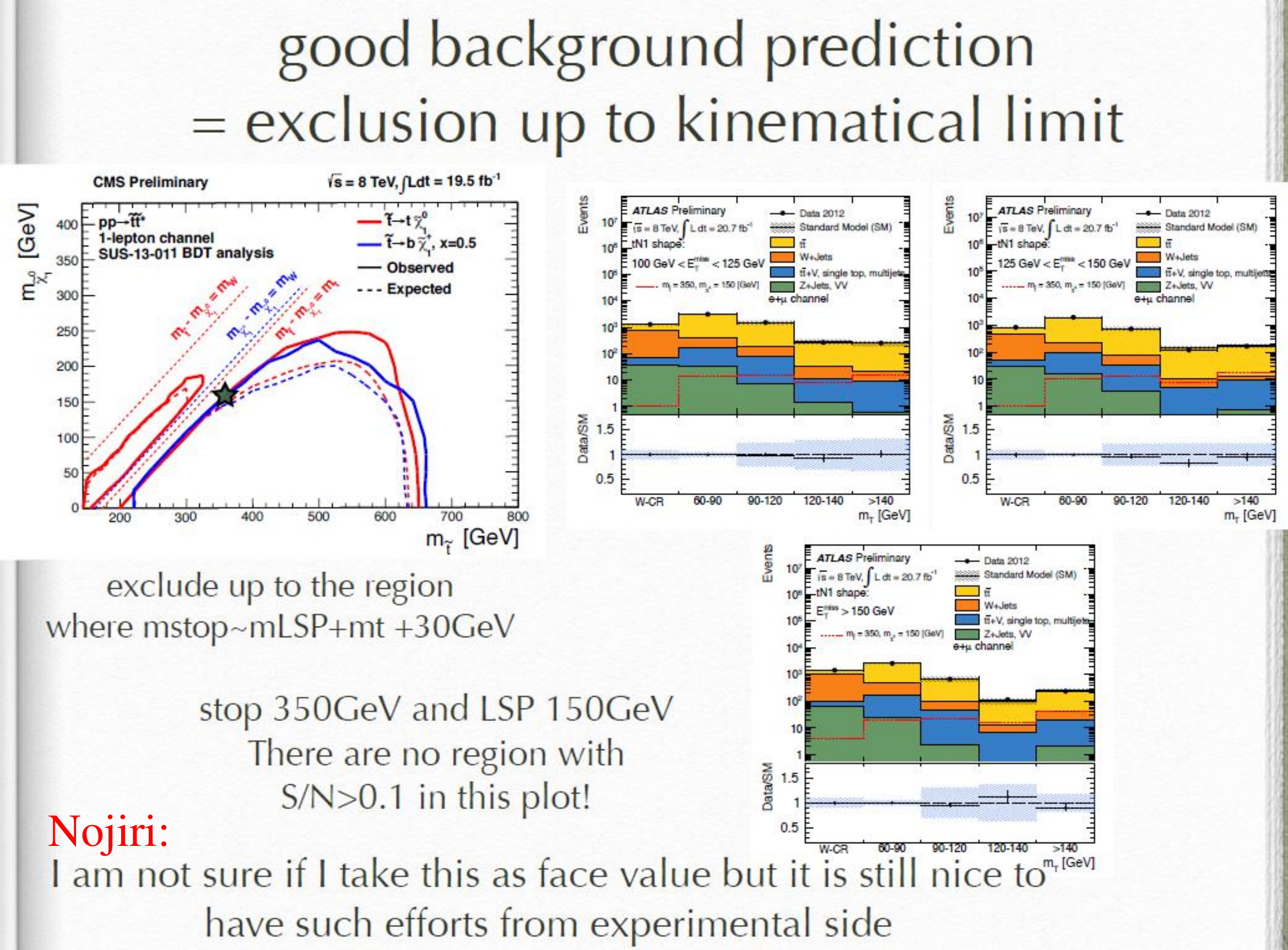}
\end{center}
\caption{\baselineskip=11pt Interplay of background estimation and exclusion limits as reviewed in Ref.~\cite{njri-lp13}.}
\label{fig13-njri}
\end{figure}

Indeed, the situation regarding the interplay between the current theoretical precision in LHC physics applications and the reach of the experimental probes of New Physics is illustrated in Figs.~\ref{fig11-njri},~\ref{fig12-njri} and ~\ref{fig13-njri} as reviewed in Ref.~\cite{njri-lp13}. In Fig.~\ref{fig11-njri}, we see that generic New Physics scenarios have the signature high $P_T$ jets, high $P_T$ leptons and $E^{\text{miss}}_T$, where in the susy case, if R-parity is violated there are additional jets and leptons instead of $E^{\text{miss}}_T$. This means that production of $W,\; Z$ and $top$ with additional jets provides significant background, so that the SM predictions for such production impact the effectiveness of the attendant experimental searches. 
Indeed, in Fig.~\ref{fig12-njri} we see the status of the comparison between some of the available theoretical predictions and the ATLAS data on on the production of $Z/\gamma^*(\rightarrow \ell^+\ell^-) + n\text{jets},\; n\ge 0$ as reviewed in Ref.~\cite{njri-lp13}. For a more inclusive observable such as the cross section for a given number of jets, the calculations featured do reasonably well relative to the respective values of $\Delta\sigma_{\text{expt}}$. For the more exclusive normalized differential distribution in $H_T$, none of the calculations featured, Alpgen~\cite{alpgen}, Sherpa~\cite{shpa}, BlackHat+Sherpa~\cite{blkht-shpa} are in good agreement with the data. Indeed, even though the BlackHat+Sherpa should have exact NLO results that Sherpa lacks, its agreement with the $H_T$ spectrum is worse in the lower and higher regions of the spectrum! One has to stress that, if the experimentalists only use the NLO results shown in Fig.~\ref{fig12-njri}, the discrepancy between the tree level and NLO results means that the error on the SM prediction therein approaches a factor of 2 at high values of $H_T$ and any limits using this spectrum would need to reflect this uncertainty. Figs.~\ref{fig12-njri} and ~\ref{fig13-njri} both then show that one needs precise background predictions to realize exclusion to the respective kinematical limit: in the latter figure, the current precision of the background predictions allows exclusion up to the region where $m_{\tilde t}=m_{\text{LSP}}+m_t + 30\text{GeV}$, where the case $m_{\tilde t}=350\text{GeV}, \; m_{\text{LSP}}=150\text{GeV}$ is illustrated.
\par
Two main observations from BSM studies here are as follows. First, conclusive results go hand-in-hand with more precise theory. Second, the transverse degrees of freedom are essential to understand with precise predictions realized on an event-by-event basis.  These observations support the need for exact resummation methods in the MC such as we have advocated in Refs.~\cite{herwiri}. Indeed,
the need for such an approach is even more acute if we consider the requirements for discovery of new heavy states: we see from the analyses above that such discovery requires strict control of the transverse degrees of freedom,
which implies exact resummation methods in the MC, for both QCD and EW higher order effects. Where do we stand on this?\par
\section{Where Do We Stand on the Required Precision Theory?}
\begin{figure}[h]
\begin{center}
\includegraphics[width=80mm]{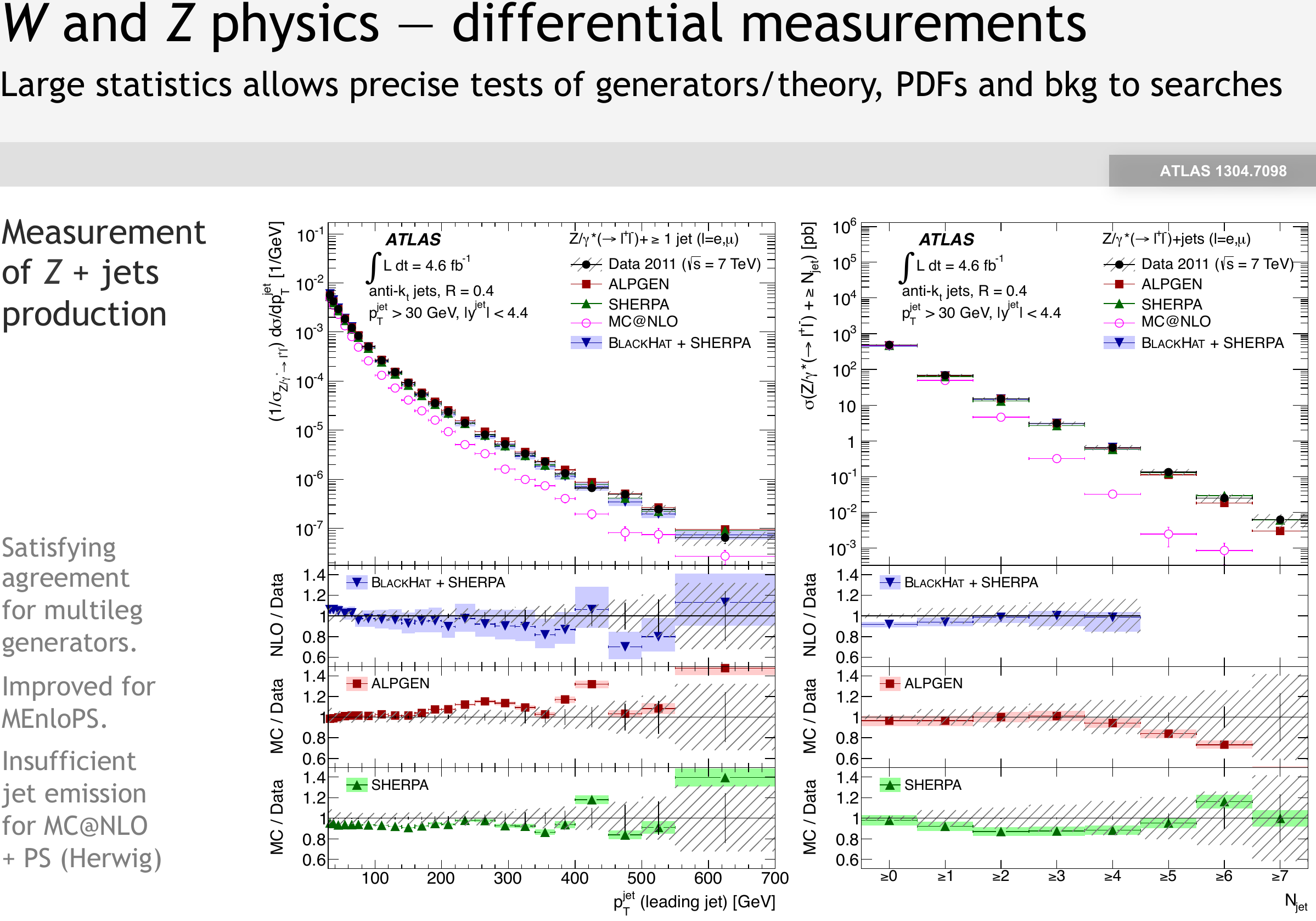}
\end{center}
\caption{\baselineskip=11pt Differential studies of W and Z physics in ATLAS as reviewed in Ref.~\cite{hoekr-lp13}.}
\label{fig14-hoekr}
\end{figure}

We use the production of $\{Z/\gamma^*,\; W^\pm\}+\text{jets}$, which was reviewed by Ref.~\cite{hoekr-lp13}, as shown here in Fig.~\ref{fig14-hoekr} to illustrate a view of the comparison of theory and experiment for SM expectations
relevant to background precision theory considerations. We do this because the large data sets for the single heavy gauge boson productions and decays to lepton pairs, which are data sets exceeding $10^7$ events in ATLAS and CMS, are not available to us at this writing. 
Focusing on the transverse momentum spectra for the leading jet, we see that, while the predictions
shown for Blackhat/Sherpa~\cite{blkht-shpa}, Sherpa~\cite{shpa} and Alpgen~\cite{alpgen} are listed as satisfactory, there is some room for improvement: indeed, if we compare Blackhat/Sherpa NLO improvement of Sherpa with Sherpa, we see that between $p^{\text{jet}}_T$ of 200GeV and 530GeV the agreement with the data is degraded by the NLO correction? In the same region, Alpgen is generally above the data, Sherpa and Blackhat/Sherpa are generally below the data, where all three sets of results are never too far from the data when the errors are taken into account. We expect that the MC@NLO\cite{mcnlo} results should agree with the low $p^{\text jet}_T$ regime(left-panel) and with the $N_{\text{jet}}\ge 0$ data point(right panel) and we see that this expectation holds. We note that all three multi-jet MC's are low at this latter right-panel point, with the NLO 
result missing it by the largest margin, $\sim 6\%$. These results show that significant improvement is needed if the analogous physics return from the analysis of the data would be desired at $300fb^{-1}$ of LHC data.\par
We note that for a large fraction of the data in Fig.~\ref{fig14-hoekr}, both $p^{\text jet}_T< 50\text{GeV}$ and $N_{\text{jet}}=0,1 $ hold. We conclude that, for normalized distributions, we must understand with good precision the
respective $\Delta\sigma_{\text{th}}$ in the regime $p^{\text jet}_T< 50\text{GeV}$ for $N_{\text{jet}}\ge 0$ -- a small error in the latter regime can result in a factor of a much larger error at large values of $p^{\text jet}_T$ and thereby change the reach of an analysis. Testing the theory predictions in this regime would be greatly aided by the release of the large ($\ge 10^7$) samples of $Z/\gamma^*$ decays to lepton pairs at ATLAS and CMS. We also stress that 
resummation~\cite{1305-0023} of the large higher order effects is essential here. Indeed, for the parton shower/ME matched exact NLO realizations in MC@NLO and Powheg~\cite{pwhg}, we have the physical theoretical precision estimate  $\Delta\sigma_{\text{phys}}\cong 10\%$. 
For the CSS~\cite{css} resummation as realized in RESBOS~\cite{resbos}, we have ~\cite{1305-0023} the estimate $\Delta\sigma_{\text{phys}}\cong {\cal O}(Q_T/Q)\cong 5\%$. For the SCT/SCET~\cite{sct,scet} resummation approach, we have~\cite{1305-0023} the estimate $\Delta\sigma_{\text{phys}}\cong \lambda = \sqrt{\Lambda/Q} \cong 5\%$. Here, we use the typical $\Lambda\cong .3\text{GeV}$, $Q= M_Z$, and $Q_T\cong 5GeV$. These estimates should be contrasted with what one expects from exact amplitude-based $QED\otimes QCD$ resummation, where~\cite{1305-0023,qced} $\Delta\sigma_{\text{phys}}\lesssim 1\%$ is possible.\par 
More precisely, standard resummations, such as that in Ref.~\cite{css} as realized in RESBOS, etc., drop terms that are not resummed as we may illustrate as follows(see Ref.~\cite{1305-0023} for more details): writing the Drell-Yan~\cite{drell-yan} formula as given in Ref.~\cite{css}(we follow the notation of this latter reference here){\small
\begin{align}
\frac{d\sigma}{dQ^2dydQ_T^2}&\sim \frac{4\pi^2\alpha^2}{9Q^2s}\Bigg\{\int \frac{d^2b}{(2\pi)^2} e^{i\vec{Q}_T\cdot\vec{b}}\sum_je_j^2\widetilde{W}_j(b^*;Q,x_A,x_B)e^{\tiny\{-\ln(Q^2/Q_0^2)g_1(b)-g_{j/A}(x_A,b)-g_{j/B}(x_B,b)\}} \nonumber\\
&\qquad\qquad\qquad\qquad\qquad +\; Y(Q_T;Q,x_A,x_B)\Bigg\},
\label{css4} 
\end{align}}
the important point is that exponentiated part of the right-hand side involves dropping terms ${\cal O}(Q_T/Q)$ to all orders in $\alpha_s$ so that, even if one knows the fixed-order based result for Y to ${\cal O}(\alpha_s^N)$ to get exactness to the latter order, one is missing terms ${\cal O}(Q_T/Q)$ in all orders
in $\alpha_s$ beyond ${\cal O}(\alpha_s^N)$.\footnote{Errors on the nonperturbative functions $g_\ell$ also yield $\Delta\sigma_{\text{phys}}$.} Corresponding arguments hold for the other approaches to resummation in Refs.~\cite{sct,scet}.\par 
What we want to call more attention to here is the following. In the usual starting point for the
fully differential representation of a hard LHC scattering process, 
\begin{equation}
d\sigma =\sum_{i,j}\int dx_1dx_2F_i(x_1)F_j(x_2)d\hat\sigma_{\text{res}}(x_1x_2s),
\label{bscfrla}
\end{equation}
where the $\{F_j\}$ and 
$d\hat\sigma_{\text{res}}$ are the respective parton densities and 
reduced hard differential cross section and where 
we use the subscript on the latter to indicate that it 
has been resummed
for all large EW and QCD higher order corrections in a manner consistent
with achieving a total precision tag of 1\% or better for the total 
theoretical precision of (\ref{bscfrla}), resummation may be carried
via resummed collinear evolution on the $\{F_j\}$ and via soft
resummation(non-collinear) on $d\hat\sigma_{\text{res}}$. 
\par
Considering first representations of resummation carried via
$d\hat\sigma_{\text{res}}$, we note the $\text{QCD}\otimes\text{QED}\equiv\text{QED}\otimes\text{QCD}$ exact amplitude-based resummation theory in Refs.~\cite{qced}, which gives{\small
\begin{eqnarray}
&d\bar\sigma_{\rm res} = \sum_{n,m}d\bar\sigma_{nm}\hspace{2.5in}\cr
& \equiv \; e^{\rm SUM_{IR}(QCED)}
   \sum_{{n,m}=0}^\infty\frac{1}{n!m!}\int\prod_{j_1=1}^n\frac{d^3k_{j_1}}{k_{j_1}} \hspace{.2in} \cr
&\qquad\qquad\qquad\prod_{j_2=1}^m\frac{d^3{k'}_{j_2}}{{k'}_{j_2}}
\int\frac{d^4y}{(2\pi)^4}e^{iy\cdot(p_1+q_1-p_2-q_2-\sum k_{j_1}-\sum {k'}_{j_2})+
D_\rQCED} \cr
&\tilde{\bar\beta}_{n,m}(k_1,\ldots,k_n;k'_1,\ldots,k'_m)\frac{d^3p_2}{p_2^{\,0}}\frac{d^3q_2}{q_2^{\,0}},
\label{subp15b}
\end{eqnarray}}\noindent
where $d\bar\sigma_{\rm res}$ is either the reduced cross section
$d\hat\sigma_{\rm res}$ or the differential rate associated to a
DGLAP-CS~\cite{dglap,cs} kernel involved in the evolution of the $\{F_j\}$ and 
where the {\em new} (YFS-style~\cite{yfs,sjbw}) {\em non-Abelian} residuals 
$\tilde{\bar\beta}_{n,m}(k_1,\ldots,k_n;k'_1,\ldots,k'_m)$ have $n$ hard gluons and $m$ hard photons and we show the generic $2f$ final state 
with momenta $p_2,\; q_2$ for
definiteness. The infrared functions ${\rm SUM_{IR}(QCED)}$, $ D_\rQCED\; $
are given in Refs.~\cite{qced,irdglap1,irdglap2}. The exactness
of the result (\ref{subp15b}) means that
its residuals $\tilde{\bar\beta}_{n,m}$ allow a rigorous parton 
shower/ME matching via their shower-subtracted 
counterparts $\hat{\tilde{\bar\beta}}_{n,m}$~\cite{qced}.\par
Indeed, focusing on the DGLAP-CS theory itself and applying 
the formula in (\ref{subp15b}) to the
calculation of the kernels, $P_{AB}$, we arrive at 
an improved IR limit of these kernels. In this IR-improved DGLAP-CS theory~\cite{irdglap1,irdglap2} large IR effects are resummed for the kernels themselves.
The resulting new resummed kernels, $P^{exp}_{AB}$~\cite{irdglap1,irdglap2}, yield a new resummed scheme for the PDF's and the reduced cross section: 
\begin{equation}
\begin{split}
F_j,\; \hat\sigma &\rightarrow F'_j,\; \hat\sigma'\; \text{for}\nonumber\\
P_{gq}(z)&\rightarrow P^{\text{exp}}_{gq}(z)=C_FF_{YFS}(\gamma_q)e^{\frac{1}{2}\delta_q}\frac{1+(1-z)^2}{z}z^{\gamma_q}, \text{etc.}.
\end{split}
\end{equation}
This new scheme gives the same value for $\sigma$ in (\ref{bscfrla}) with improved MC stability
as discussed in Ref.~\cite{herwiri}. Here, the YFS~\cite{yfs} infrared factor 
is given by $F_{YFS}(a)=e^{-C_Ea}/\Gamma(1+a)$ where $C_E$ is Euler's constant
and we refer the reader to Ref.~\cite{irdglap1,irdglap2} for the definition of the infrared exponents $\gamma_q,\; \delta_q$ as well as for the complete
set of equations for the new $P^{exp}_{AB}$. $C_F$ is the quadratic Casimir invariant for the quark color representation.\par
For reference, to show how we make
contact between the $\hat{\tilde{\bar\beta}}_{n,m}$ and the
differential distributions in MC@NLO we proceed as
follows. We represent the MC@NLO differential cross section 
via~\cite{mcnlo} 
\begin{equation}
\begin{split}
d\sigma_{MC@NLO}&=\left[B+V+\int(R_{MC}-C)d\Phi_R\right]d\Phi_B[\Delta_{MC}(0)+\int(R_{MC}/B)\Delta_{MC}(k_T)d\Phi_R]\nonumber\\
&\qquad\qquad +(R-R_{MC})\Delta_{MC}(k_T)d\Phi_Bd\Phi_R
\label{mcatnlo1}
\end{split}
\end{equation}
where $B$ is Born distribution, $V$ is the regularized virtual contribution,
$C$ is the corresponding counter-term required at exact NLO, $R$ is the respective
exact real emission distribution for exact NLO, $R_{MC}=R_{MC}(P_{AB})$ is the parton shower real emission distribution
so that the Sudakov form factor is 
$$\Delta_{MC}(p_T)=e^{[-\int d\Phi_R \frac{R_{MC}(\Phi_B,\Phi_R)}{B}\theta(k_T(\Phi_B,\Phi_R)-p_T)]}$$,
where as usual it describes the respective no-emission probability.
The respective Born and real emission differential phase spaces are denoted by $d\Phi_A, \; A=B,\; R$, respectively.
We may note further that the representation of the differential distribution
for MC@NLO in (\ref{mcatnlo1}) is an explicit realization of the compensation 
between real and virtual divergent soft effects discussed in the 
Appendices of Refs.~\cite{irdglap1,irdglap2} in establishing the validity of 
(\ref{subp15b}) for QCD -- all of the terms on the RHS of (\ref{mcatnlo1}) are 
infrared finite. Indeed,
from comparison with (\ref{subp15b}) restricted to its QCD aspect we get the identifications, accurate to ${\cal O}(\alpha_s)$,
\begin{equation}
\begin{split}
\frac{1}{2}\hat{\tilde{\bar\beta}}_{0,0}&= \bar{B}+(\bar{B}/\Delta_{MC}(0))\int(R_{MC}/B)\Delta_{MC}(k_T)d\Phi_R\\
\frac{1}{2}\hat{\tilde{\bar\beta}}_{1,0}&= R-R_{MC}-B\tilde{S}_{QCD}
\label{eq-mcnlo}
\end{split}
\end{equation}
where we defined~\cite{mcnlo} $$\bar{B}=B(1-2\alpha_s\Re{B_{QCD}})+V+\int(R_{MC}-C)d\Phi_R$$ and we understand here
that the DGLAP-CS kernels in $R_{MC}$ are to be taken as the IR-improved ones
as we exhibit below~\cite{irdglap1,irdglap2}. 
Here we have written the QCD virtual and real infrared functions
$B_{QCD}$ and $\tilde{S}_{QCD}$ respectively without the superscript $nls$
for simplicity of notation and they are understood to be DGLAP-CS synthesized as explained in Refs.~\cite{qced,irdglap1,irdglap2} so that we
avoid doubling counting of effects. We also re-emphasize that we do not drop
any effects here in (\ref{eq-mcnlo}). We observe further that, in view of 
(\ref{eq-mcnlo}), 
the way to the extension of frameworks such as MC@NLO to exact higher
orders in $\{\alpha_s,\;\alpha\}$ is therefore open via our $\hat{\tilde{\bar\beta}}_{n,m}$
and will be taken up elsewhere~\cite{elswh}.\par
\begin{figure}[h]
\begin{center}
\includegraphics[width=80mm]{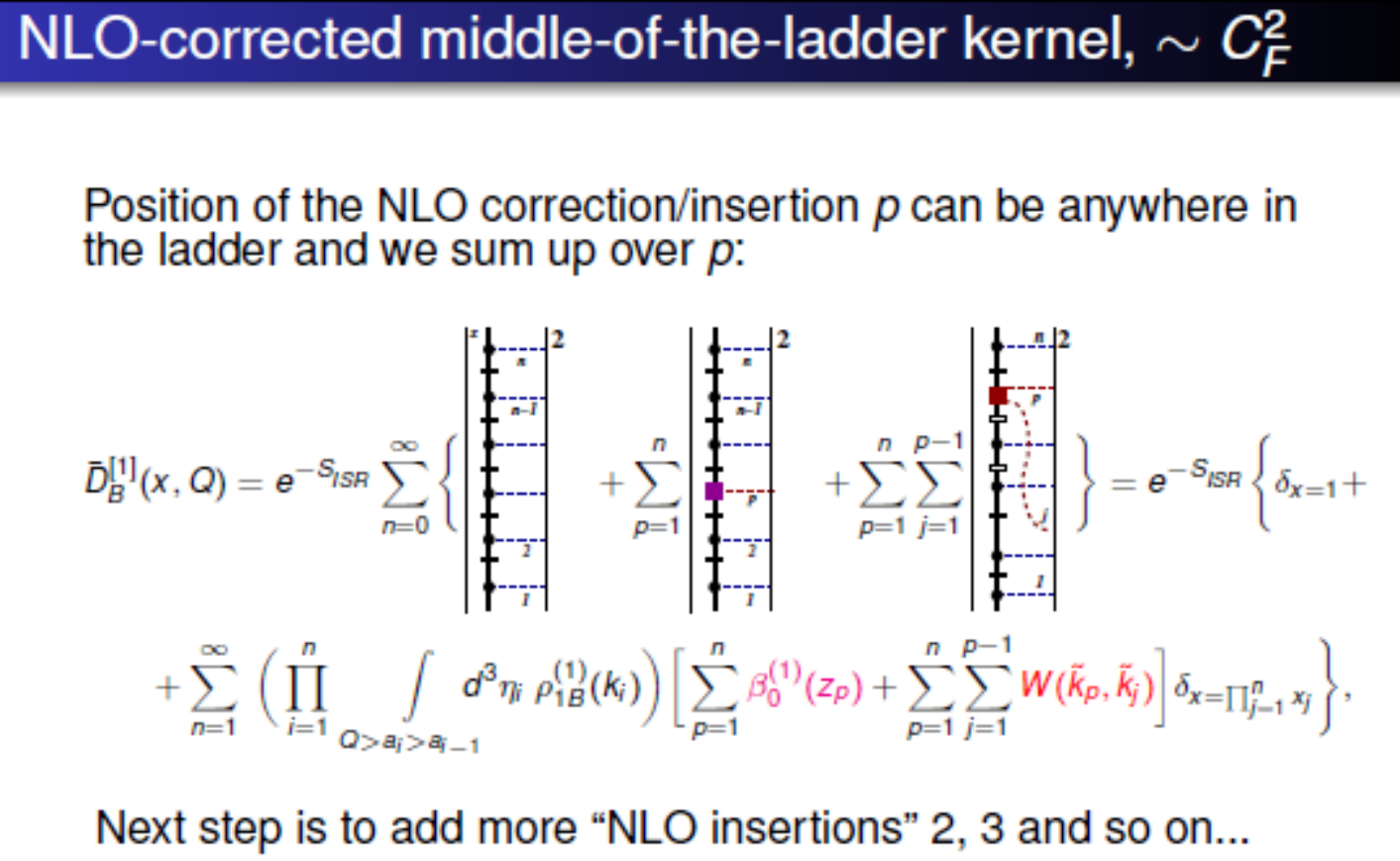}
\end{center}
\caption{\baselineskip=11pt Proof of concept for the non-singlet NLO DGLAP evolution from Ref.~\cite{jadskrpk}.}
\label{fig15-jadskrpk}
\end{figure}
\begin{figure}[h]
\begin{center}
\includegraphics[width=80mm]{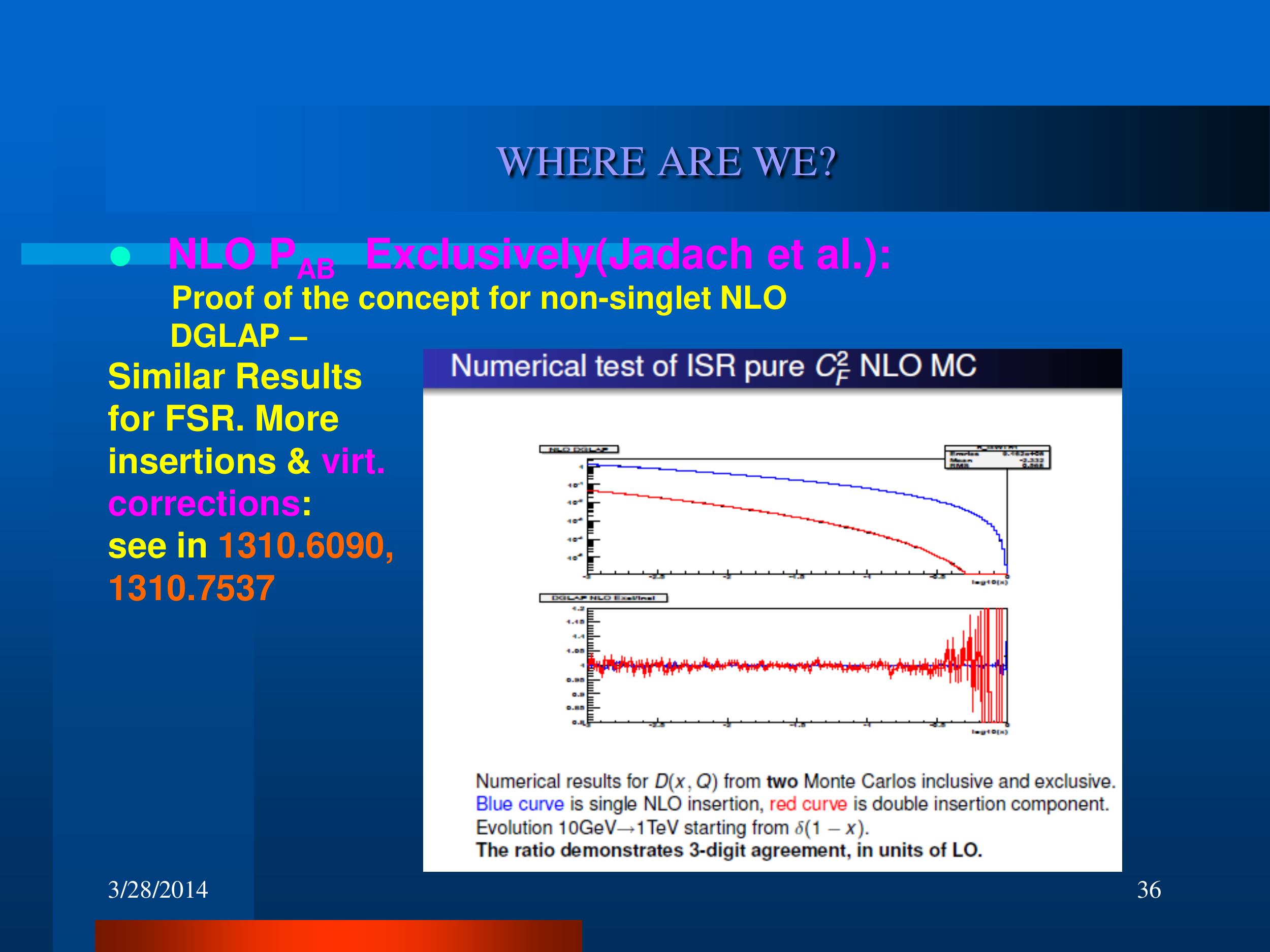}
\end{center}
\caption{\baselineskip=11pt Numerical tests with one and two NLO insertions in the NLO DGLAP evolution from Ref.~\cite{jadskrpk}.}
\label{fig16-jadskrpk}
\end{figure}

Turning next to the representations of resummation carried by the $\{F_J\}$
we note the NLO exclusive improvement of the $P_{AB}$ in the parton shower evolution as developed in Ref.~\cite{jadskrpk}. As we see in Figs.~\ref{fig15-jadskrpk} and ~\ref{fig16-jadskrpk}, the proof of concept for the non-singlet NLO DGLAP evolution has been established with successful numerical tests of the ISR pure $C_F^2$ NLO MC. 
Similar results have been obtained for the FSR and results with more than two NLO insertions and with virtual corrections are also now available~\cite{jadskrpk}. The way is therefore open for the complete NLO in the parton shower itself.
\par
\begin{figure}[h]
\begin{center}
\includegraphics[width=80mm]{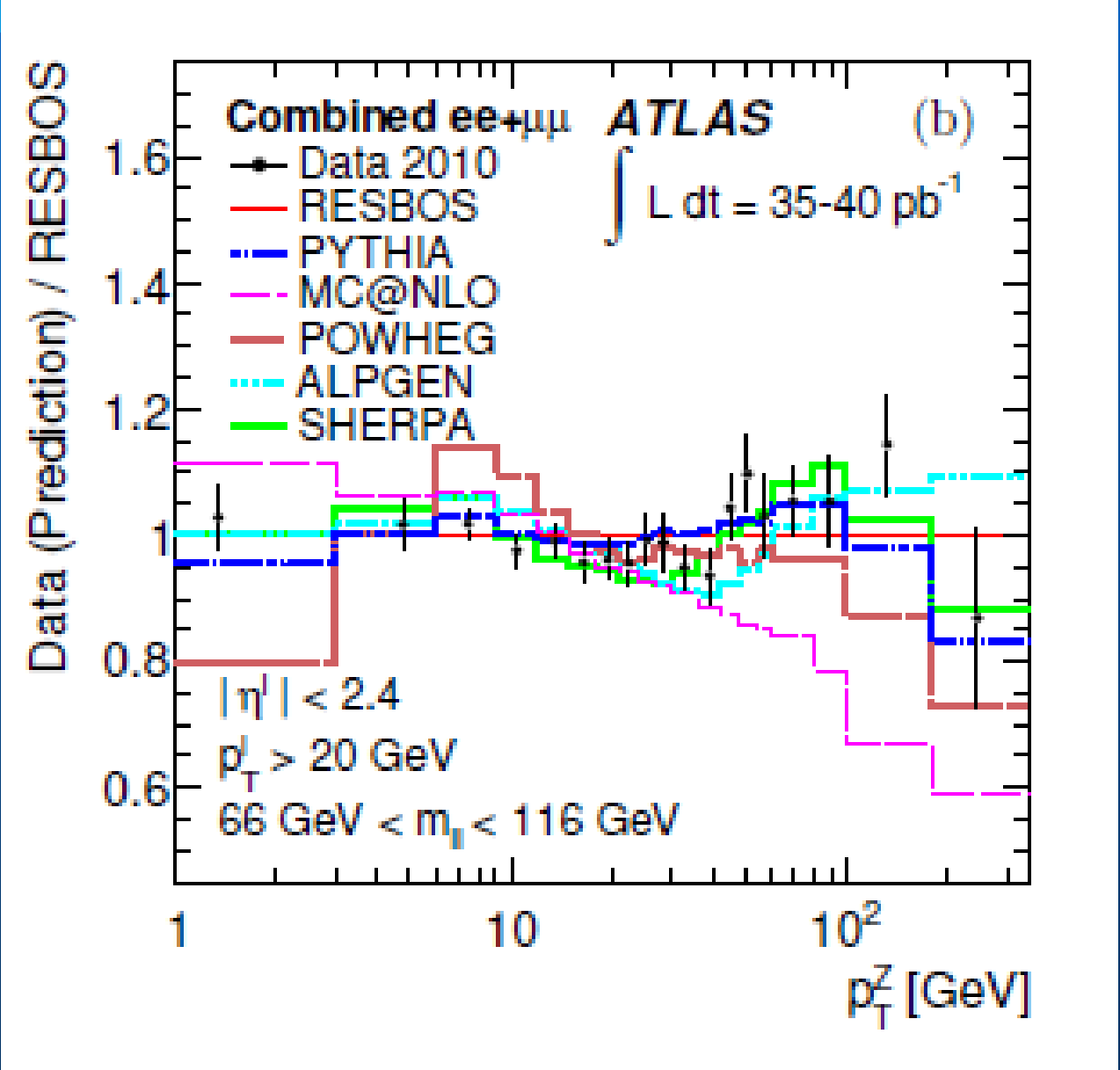}
\end{center}
\caption{\baselineskip=11pt Comparisons of some theoretical predictions with the ATLAS  $Z\gamma^*$ $p_T$ spectrum in single $Z\gamma^*$ production with decay to lepton pairs as given in the first reference in Refs.~\cite{atlaspt-phi-eta-str}.}
\label{fig17-atlas}
\end{figure}
\begin{figure}[h]
\begin{center}
\includegraphics[width=80mm]{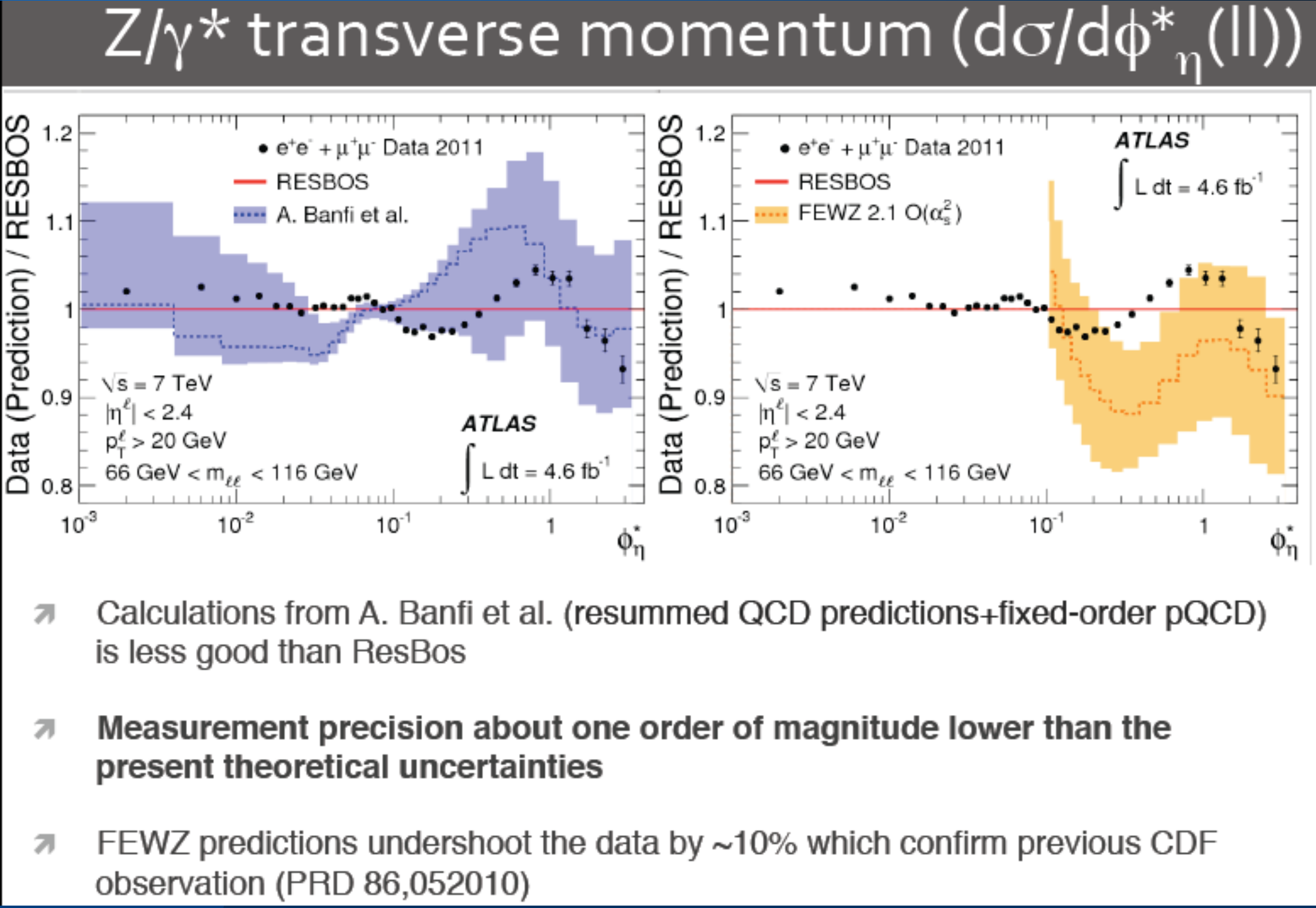}
\end{center}
\caption{\baselineskip=11pt Comparisons of some theoretical predictions with the ATLAS  $Z\gamma^*$ $\phi^*_\eta$ spectrum in single $Z\gamma^*$ production with decay to lepton pairs as given in the second reference in Refs.~\cite{atlaspt-phi-eta-str}.}
\label{fig18-atlas}
\end{figure}

Given that we have approaches that in principle can reach the sub-1\% precision theoretical precision tag for LHC physics process, what is the current state of affairs in the comparison between the theoretical predictions in the refereed literature and the LHC data? We will illustrate the situation first with the results from ATLAS in Figs.~\ref{fig17-atlas} and ~\ref{fig18-atlas}
from Refs.~\cite{atlaspt-phi-eta-str}. In the former figure, some calculations available in the literature are compared to the combined
ATLAS $e\bar{e}$ and $\mu\bar{\mu}$ data from 2010 for the differential 
$p_T$ spectrum 
in single $Z/\gamma^*$
production and decay to lepton pairs and in latter figure another set of calculations available in the literature are compared to the ATLAS data for the 
differential $\phi^*_\eta$ spectrum, where $\phi^*_\eta=\tan(\frac{1}{2}(\pi-\Delta\phi))\sin\theta^*$, with $\sin\theta^*=1/\cosh(\Delta\eta/2)$.
Here, $\Delta\phi,\; \Delta\eta$ are the respective differences in the 
the azimuths and pseudo-rapidities of the two attendant leptons. 
We see in both cases
that there is considerable need for improvement if we want to have 
the same or a better reach in physics for $300fb^{-1}$ of LHC data.
Indeed, given that none of the calculations actually ``looks'' like the data
for all values of the observables plotted,
one could even question what reach in physics 
these data allow now
at the LHC. Indeed, since there are other calculations available in the literature that are excluded from the comparisons in Figs.~\ref{fig17-atlas} and ~\ref{fig18-atlas}, this leaves the question even more unsettled.\par
\begin{figure}[ht]
\begin{center}
\setlength{\unitlength}{0.1mm}
\begin{picture}(1600, 930)
\put( 370, 770){\makebox(0,0)[cb]{\bf (a)} }
\put(1240, 770){\makebox(0,0)[cb]{\bf (b)} }
\put(   -50, 0){\makebox(0,0)[lb]{\includegraphics[width=80mm]{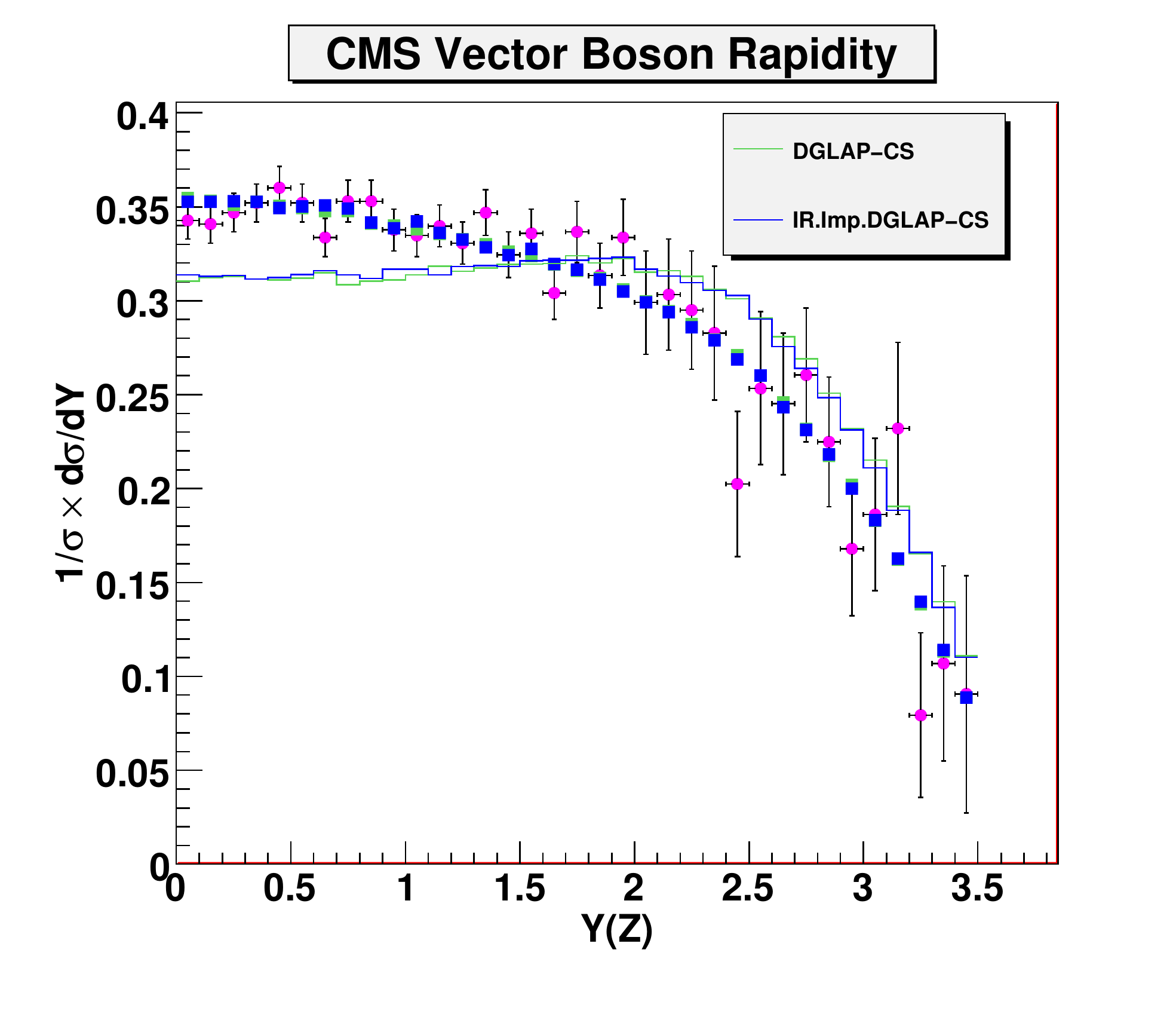}}}
\put( 830, 0){\makebox(0,0)[lb]{\includegraphics[width=80mm]{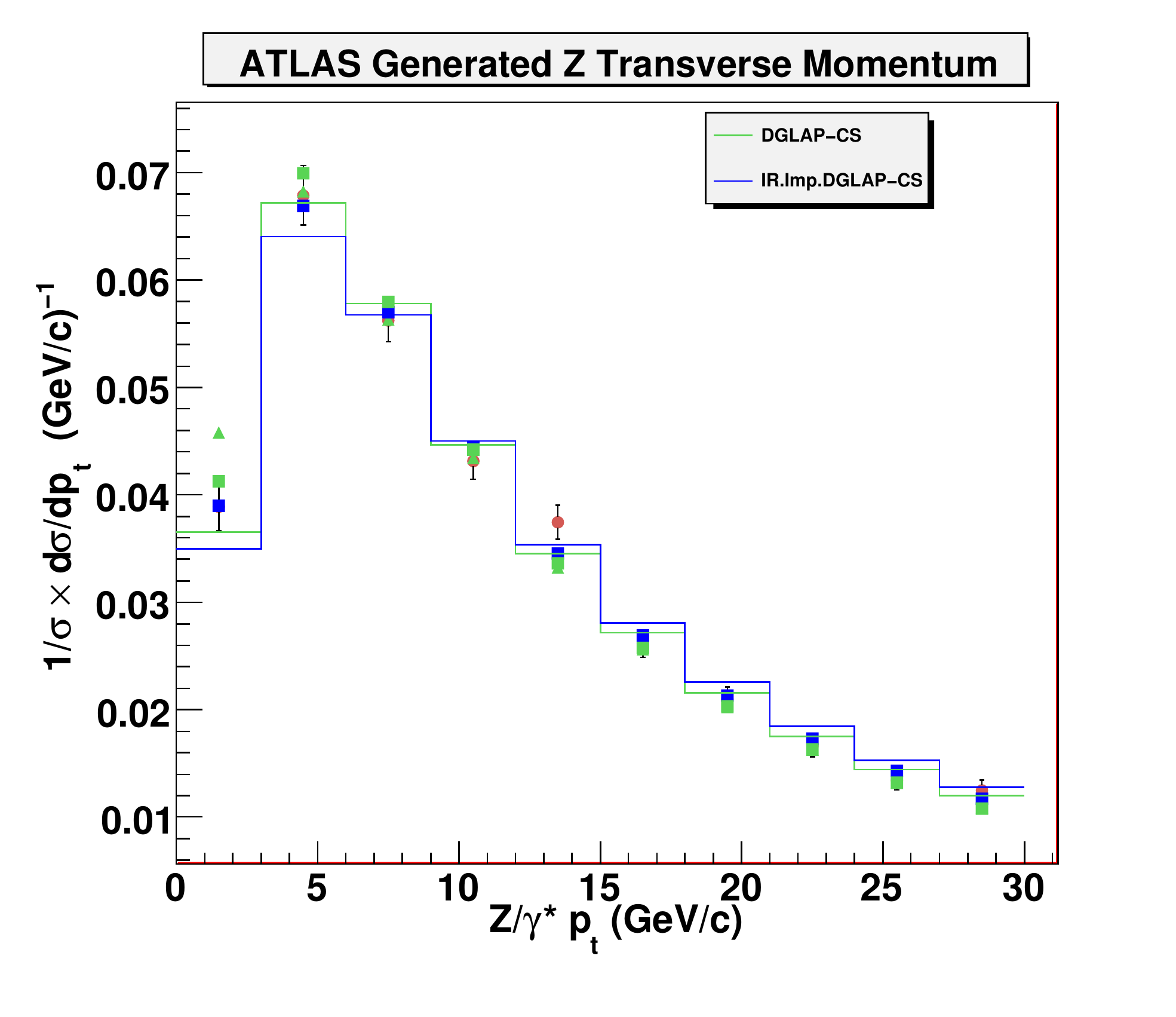}}}
\end{picture}
\end{center}
\caption{\baselineskip=8pt Comparison with LHC data: (a), CMS rapidity data on
($Z/\gamma^*$) production to $e^+e^-,\;\mu^+\mu^-$ pairs, the circular dots are the data, the green(blue) lines are HERWIG6.510(HERWIRI1.031); 
(b), ATLAS $p_T$ spectrum data on ($Z/\gamma^*$) production to (bare) $e^+e^-$ pairs,
the circular dots are the data, the blue(green) lines are HERWIRI1.031(HERWIG6.510). In both (a) and (b) the blue(green) squares are MC@NLO/HERWIRI1.031(HERWIG6.510($\rm{PTRMS}=2.2$GeV)). In (b), the green triangles are MC@NLO/HERWIG6.510($\rm{PTRMS}=$0). These are otherwise untuned theoretical results. 
}
\label{fig19-hwriri}
\end{figure}

In Refs.~\cite{herwiri}, we have realized the new IR-improved DGLAP-CS theory in
the Herwig6.5 environment with the new MC Herwiri1.031. We show in Figs.~\ref{fig19-hwriri} comparison between the predictions for the $p_T$ and rapidity spectra
for the IR-improved MC Herwiri1.031 and the predictions for the well-known Herwig.5 MC without our IR-improvement in its parton showers, where in the case of Herwig6.5, we show spectra both for an intrinsic Gaussian $p_T$ with rms value 2.2GeV and without such intrinsic $p_T$ for the proton constituents. 
For the unimproved case, the comparisons with the ATLAS $p_T$ data~\cite{atlaspt-phi-eta-str} suggest that we need the Gaussian (intrinsic) $p_T$ with rms value $\text{\rm PTRMS}\cong 2.2\text{GeV}$ to get a good fit to both of the spectra, wheres for the IR-improved case(Herwiri1.031), we get good fits to both sets of spectra without the need of such a hard intrinsic
$p_T$. Both the IR-improved and the IR-unimproved MC's need the MC@NLO exact ${\cal O}(\alpha_s)$ correction to fit the CMS rapidity data~\cite{cms-rap} shown.
The respective sets of $\chi^2/d.o.f's$ are $\{1.37,0.70\}$, $\{0.72,0.72\}$,
and $\{2.23,0.70\}$ for the MC@NLO/Herwig6.510($\text{\rm PTRMS}=2.2\text{GeV}$), MC@NLO/Herwiri1.031, and 
MC@NLO/Herwig6.510($\text{\rm PTRMS}=0.0$) predictions for the $p_T$ and rapidity spectra.\par
\begin{figure}[ht]
\begin{center}
\includegraphics[width=80mm]{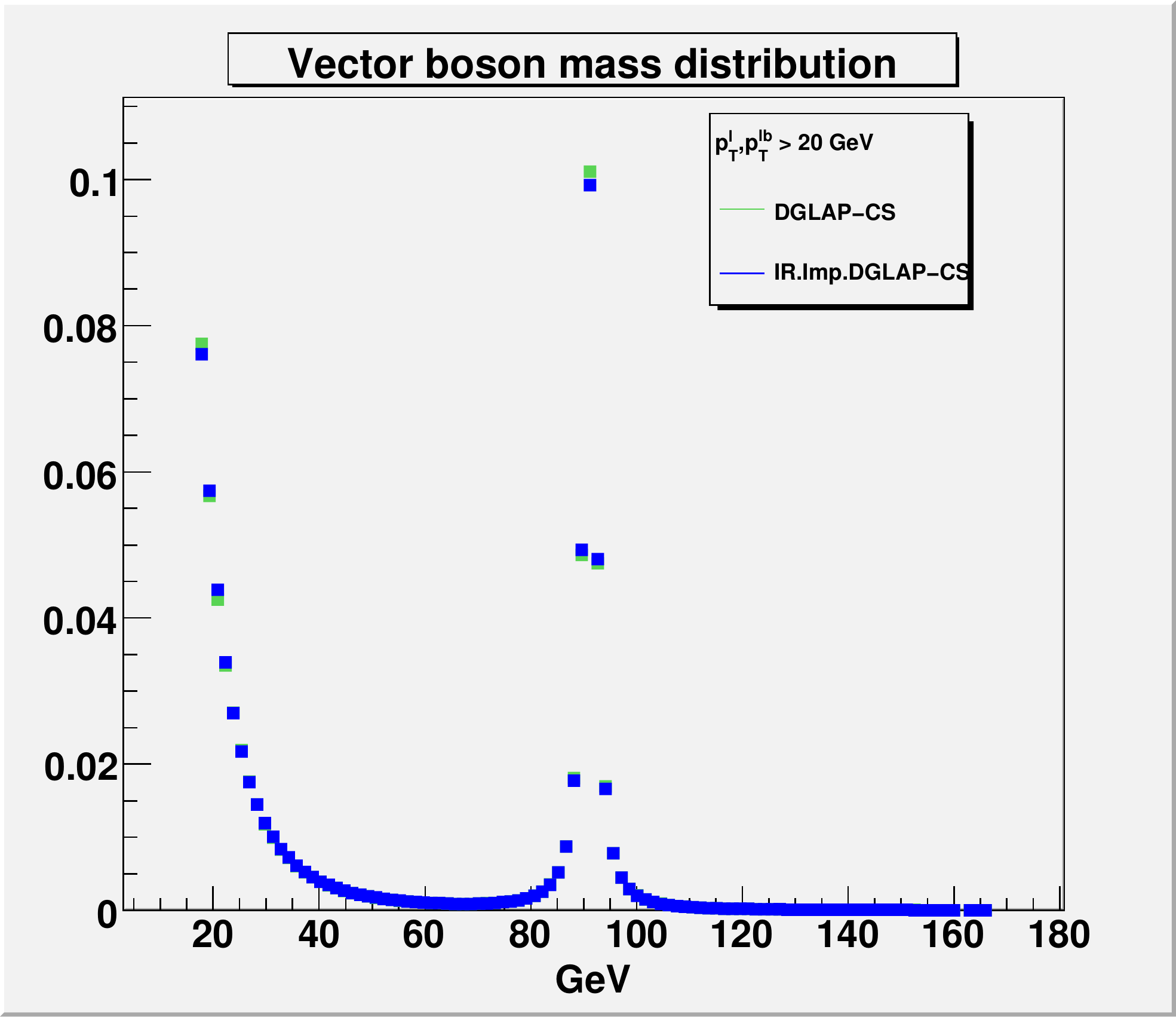}
\end{center}
\caption{Normalized vector boson mass spectrum at the LHC for $p_T(\text{lepton}) >20$ GeV.}
\label{fig20-hwriri}
\end{figure}
\begin{figure}[ht]
\begin{center}
\includegraphics[width=80mm]{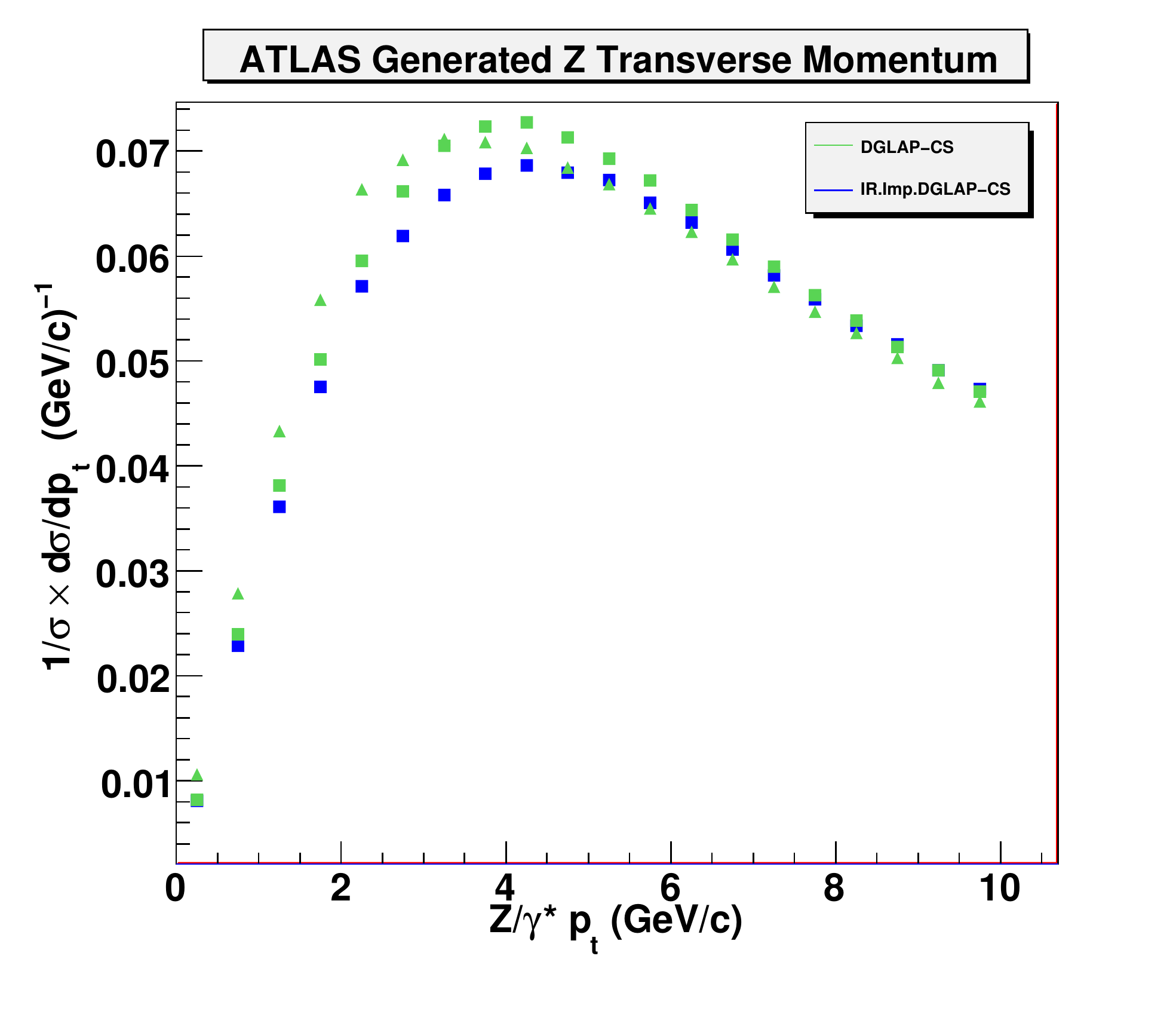}
\end{center}
\caption{Normalized vector boson $p_T$ spectrum at the LHC 
for the ATLAS cuts as exhibited in Fig.~\protect{\ref{fig19-hwriri}} 
for the same conventions on the notation for the theoretical results
with the vector boson $p_T < 10$ GeV to illustrate the differences between
the three predictions.}
\label{fig21-hwriri}
\end{figure}

Which of the predictions illustrated in Fig.~\ref{fig19-hwriri} is better for precision QCD theory? We note that precocious Bjorken scaling in the SLAC-MIT experiments~\cite{bj1,taylor}, where scaling occurs already at $Q^2\cong 1_+\text{GeV}^2$, implies that $\text{\rm PTRMS}^2$ should be small compared to $1_+ \text{GeV}^2$. See the models of the proton wave function in Refs.~\cite{protonwfn}, where in all cases $\text{\rm PTRMS}^2<< 1_+ \text{GeV}^2 $. This favors the
IR-improved approach in Herwiri1.031.
Moreover, the first principles approach in Herwiri1.031 is not subject to arbitrary functional variation to determine its $\Delta\sigma_{\text th}$. Experimentally, in principle, the events in the two cases MC@NLO/Herwig6.510($\text{\rm PTRMS}=2.2\text{GeV}$) and  MC@NLO/Herwiri1.031 should look different in terms of the properties of the rest of the particles in the events -- this is under study~\cite{elswh}. Here we show already the results in
in Figs.~\ref{fig20-hwriri} and ~\ref{fig21-hwriri}, which show that
the two MC's predict a 2.2\% difference in the Z-peak and that, if we 
make a finer binning of the $p_T$ spectra, 0.5GeV/bin instead of 3GeV/bin, 
we can distinguish
all three cases, MC@NLO/Herwig6.510($\text{\rm PTRMS}=2.2\text{GeV}$), 
MC@NLO/Herwiri1.031, and 
MC@NLO/Herwig6.510($\text{\rm PTRMS}=0.0$), with the type of precision data 
at the LHC ($\ge 10^7 Z/\gamma^*$ decays to lepton pairs).
We await these exciting data.
We stress that IR-improved DGLAP-CS theory increases the definiteness of precision determination of NLO parton shower MC's and improves such. More potential checks against experiment of the new IR-improved MC Herwiri1.031 are being pursued.
We note that realizations of the new IR-improved showers are in progress for
Herwig++~\cite{hwg++,mke1}, Pythia8~\cite{pythia8,jc-ps-ts} and Sherpa~\cite{janw}.
\par 
In the near future, in addition to more specific tests of observables such as 
$\phi^*_\eta$, with and without $p_T$ cuts on the respective $Z\gamma^*$, as well as the analysis s of the large and hopefully soon to be released $> 10^7$
lepton pairs data samples form ATLAS and CMS, we are also pursuing the
version Herwiri2.0~\cite{herwiri2} in which the CEEX realization of higher order EW corrections in \KKMC~\cite{kkmc1} is realized in the Herwig6.5 environment
and the direct application of \KKMC to LHC processes with the release recently of \KKMC 4.22~\cite{kkmc2} in which the incoming beam choices are extended to
$q\bar{q}, \mu\bar{mu},\tau\bar{\tau}, \nu_\ell\bar{\nu_\ell}, \; q=u,d,s,c,b,t,
\; \ell=e,\mu,\tau$. Finally, we note that we have in mind as well the 
development of MC@NNLO or its equivalent, as that is the requirement for
the sub-1\% precision tag that will be needed to exploit fully the $300fb^{-1}$ 
LHC data sets.\par
We show in Tab.~\ref{tab:table2-kkmc} and
Fig.~\ref{fig23-kkmc}  
the respective illustrative results 
for $d\bar{d}\rightarrow \mu\bar{\mu}$ at $189$GeV
for its $v$-spectrum\footnote{Here, $v=1-s'/s$ in the usual notation.} and for its physical precision test
and in Fig.~\ref{fig24-kkmc} the illustrative results for $pp\rightarrow u\bar{u}\rightarrow \ell\bar{\ell}+n\gamma$ where the proton PDF's for $u$ and $\bar{u}$ are replacing the beamsstrahlung distributions in \KKMC.
\def\Energy{189GeV} 
\def\Process{$d \bar{d} \to \mu^-\mu^+$}
\begin{table}[ht]
\centering
\includegraphics[width=90mm,height=60mm]{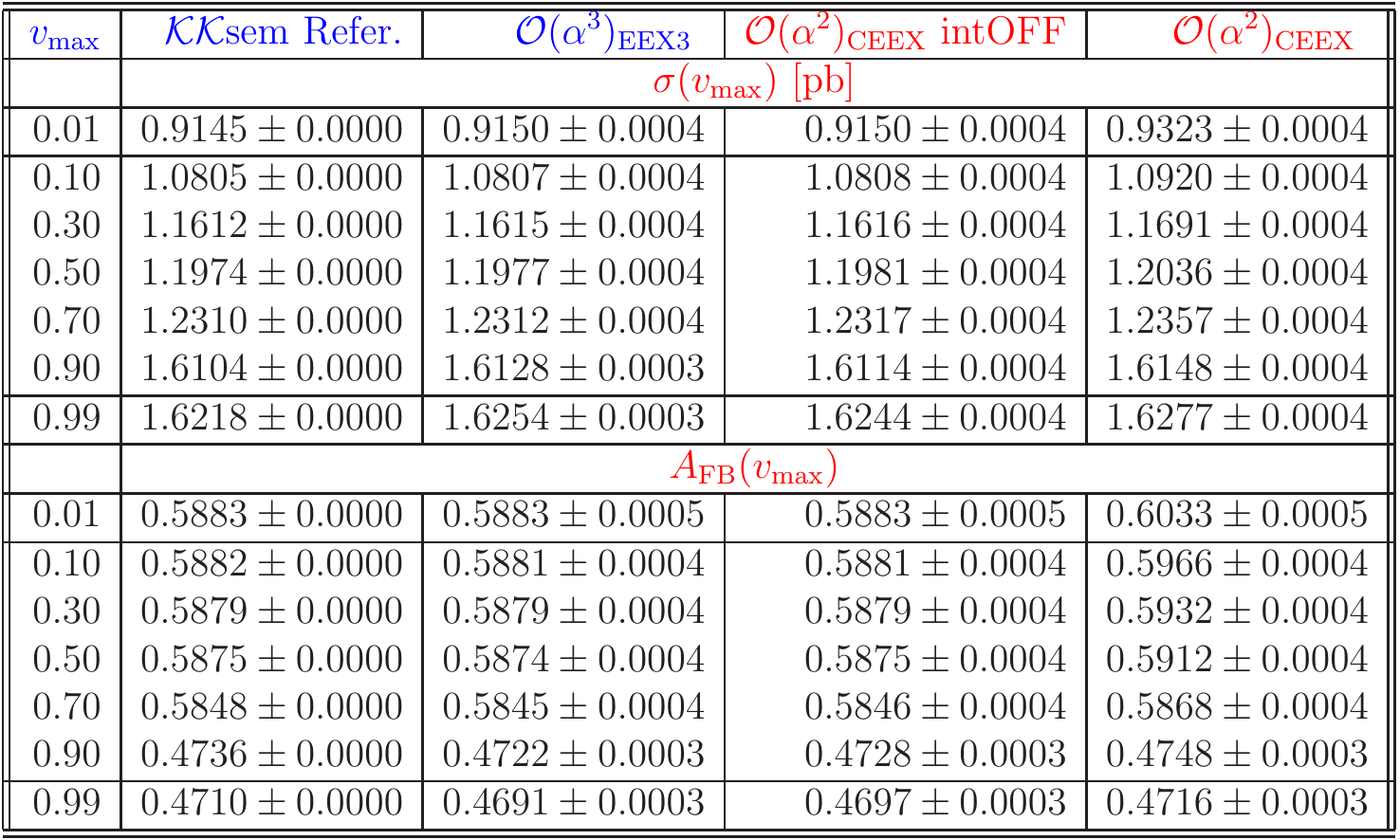}
\caption{
 Study of total cross section $\sigma(v_{\max})$ 
 and charge asymmetry $A_{\rm FB}(v_{\max})$,
 \Process, at $\sqrt{s}$~=\Energy.
 See the text for definition of
 the energy cut $v_{\max}$, see Ref.~\cite{kkmc2} for the scattering angle and M.E. type definitions. 
}
\label{tab:table2-kkmc}
\end{table}
\begin{figure}
\centering
\includegraphics[width=71mm,height=40mm]{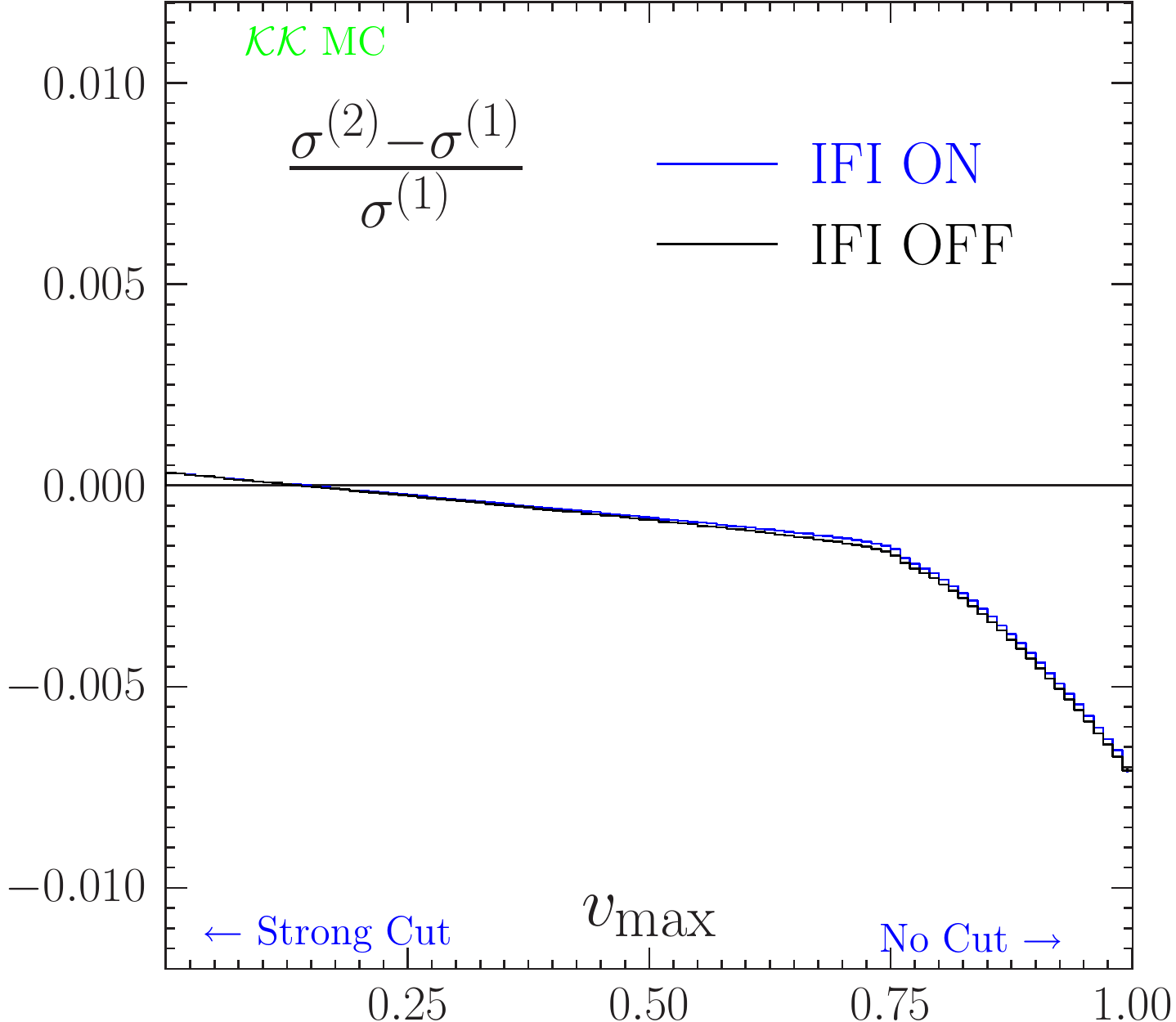}
\includegraphics[width=71mm,height=40mm]{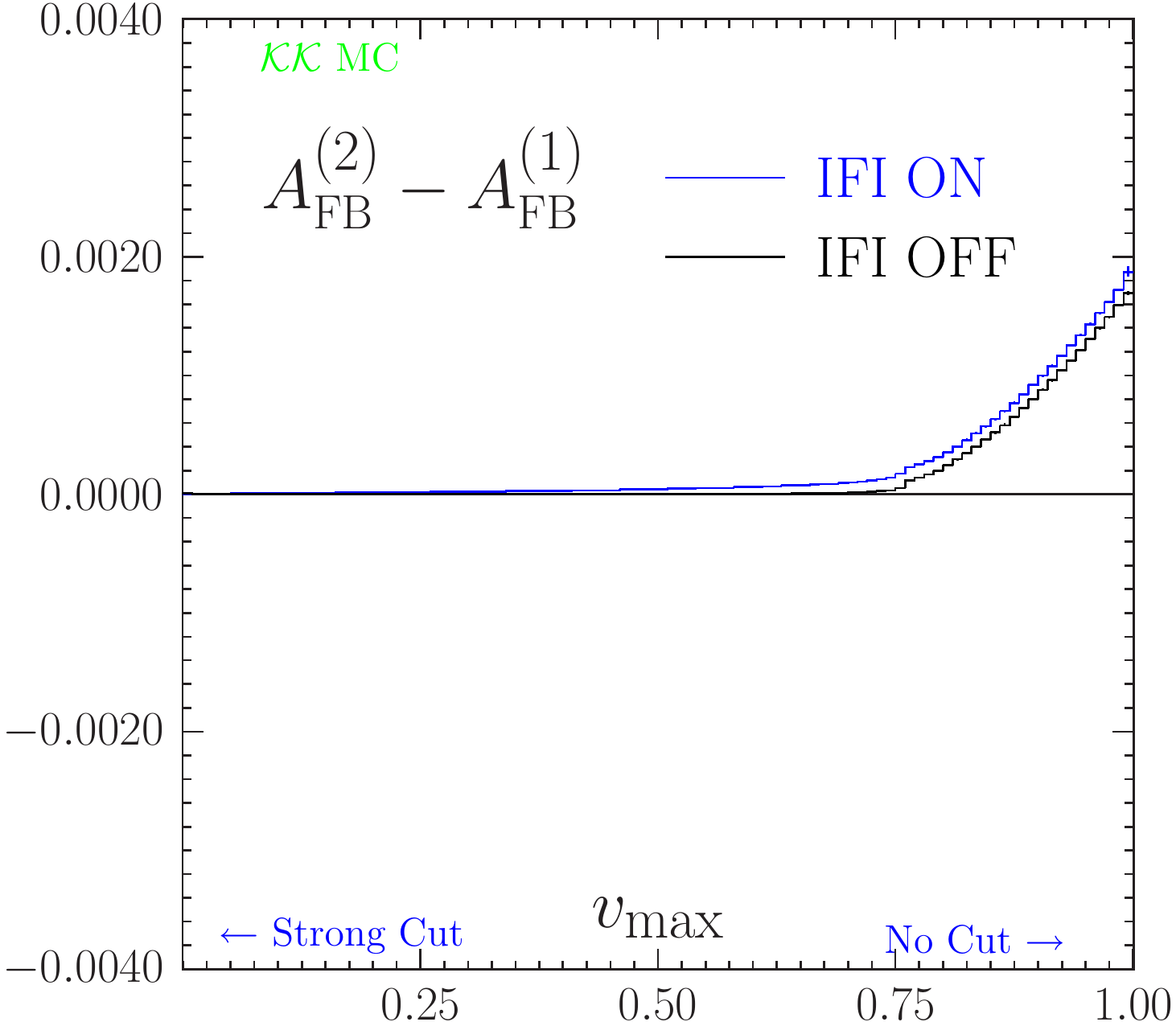}
\caption{
Physical precision of CEEX ISR matrix element
for \Process at $\sqrt{s}=$\Energy.
See table \ref{tab:table2-kkmc} for definition
of cut-offs.
}
\label{fig23-kkmc}
\end{figure}
\begin{figure}
\begin{center}
\setlength{\unitlength}{0.1mm}
\begin{picture}(1600, 930)
\put(   -50, 0){\makebox(0,0)[lb]{\includegraphics[width=75mm]{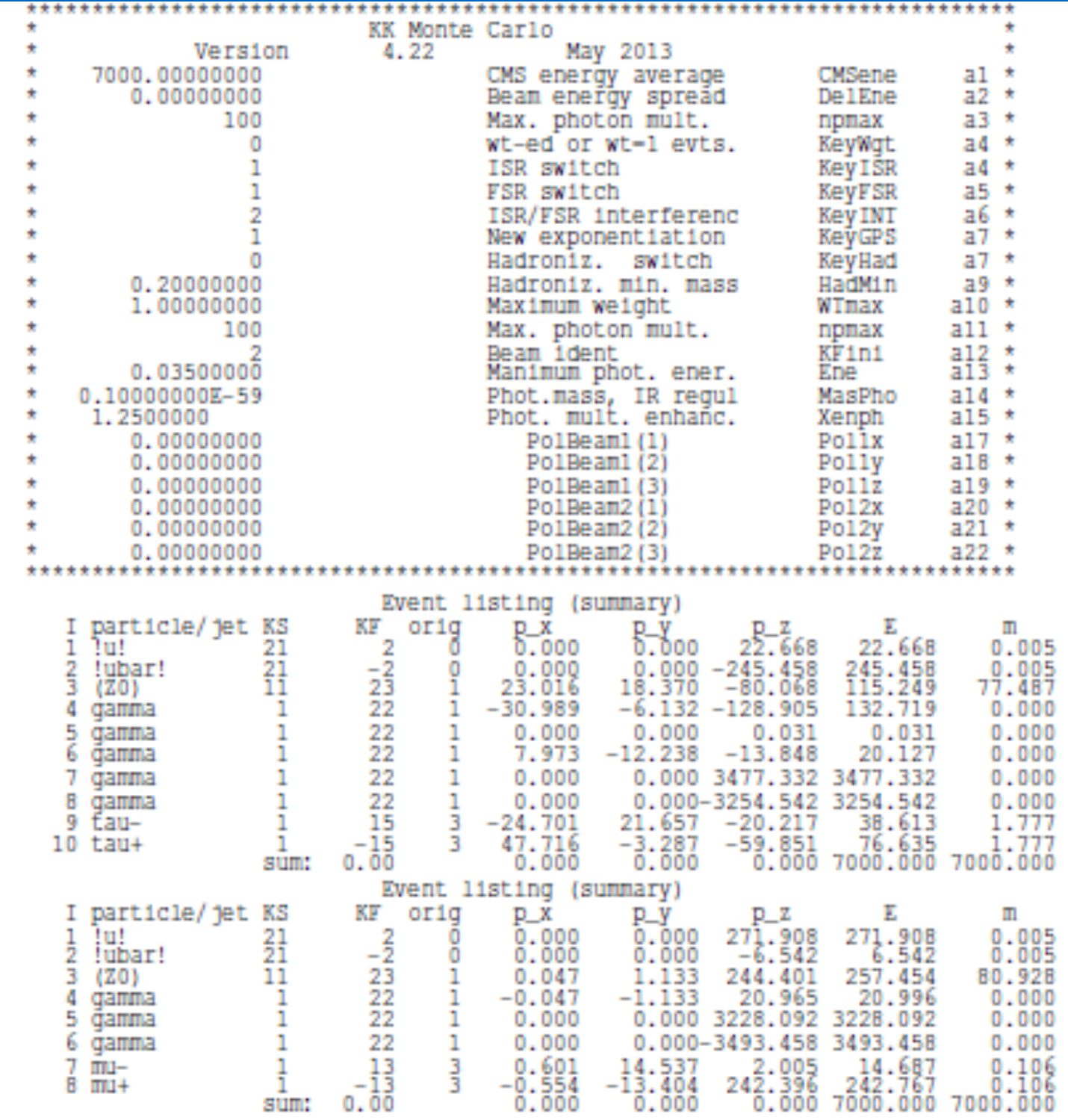}}}
\put( 750,530){\makebox(0,0)[lb]{\includegraphics[width=75mm]{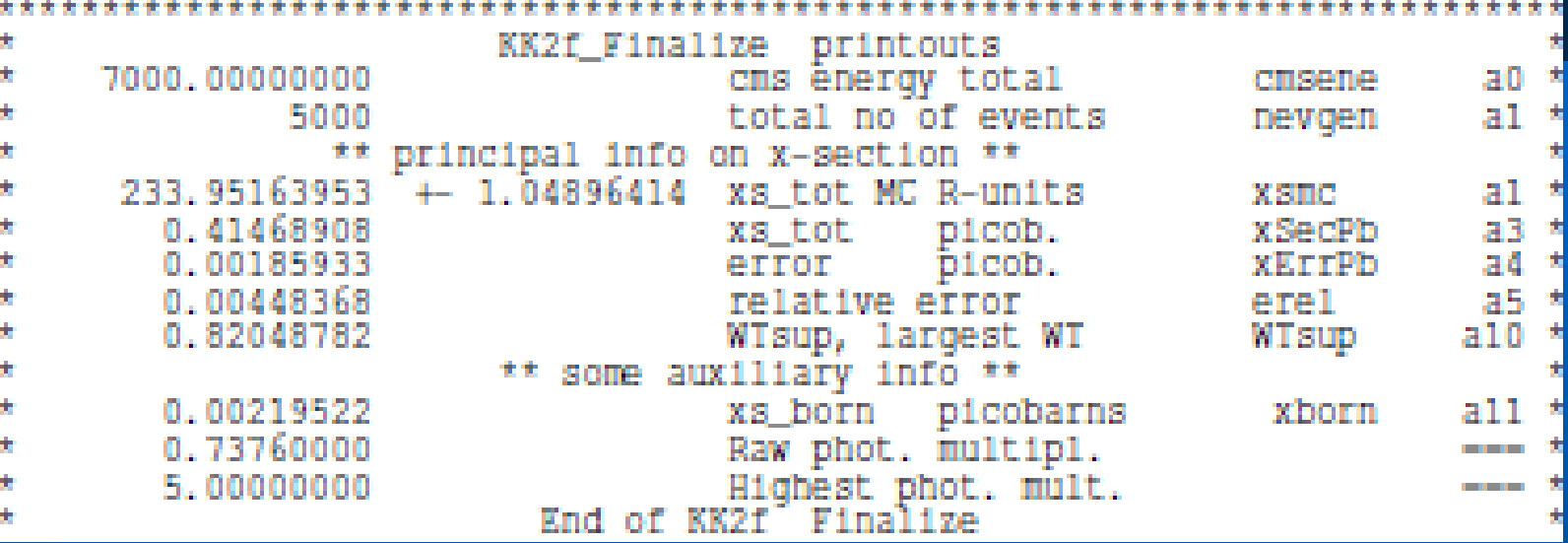}}}
\end{picture}
\end{center}
\caption{
Sample output from \KKMC 4.22 for $pp\to u\bar{u} \to l^- l^+ +n\gamma$.
}
\label{fig24-kkmc}
\end{figure}

Take note 
that the results show the value $\Delta\sigma_{\text{th}}\cong 0.1\%$ for the $d\bar{d}$ process
at the typical energy cut near $v\cong 0.6$ and show effects of non-zero $p_T$ in the multiple photon radiation at the LHC can be important. These matters are under study accordingly.\par
\section{OUTLOOK}
We note that there are other efforts than those we have mentioned to 
improve resummation in progress. In the EW collinear regime we call 
attention to Refs.~\cite{barze}, in QCD we note the new SCET approach 
of Ref.~\cite{newscet}, and the new SCET based MC's in 
Refs.~\cite{scetmc}, etc. 
There are new NLO and new NNLO results: multi-leg NLO~\cite{mltilg-nlo}, 
$t\bar{t}$ at NNLO~\cite{t-t-barnnlo}, etc. What we can say is that the 
full exploitation of the LHC discovery potential with $300fb^{-1}$ of 
data will need all such efforts in view of and in conjunction with 
what we have discussed above. In closing,
we thank Profs. S. Jadach and S. Yost and Dr. S. Majhi for useful discussions 
and we also thank Prof. Ignatios Antoniadis for the support and kind 
hospitality of the CERN TH Unit while part of this work was completed.\par

\end{document}